\documentclass[superscriptaddress,
twocolumn,
amsmath,amssymb]{revtex4}
\usepackage{epsfig}
\usepackage{bm}
\usepackage{times}
\usepackage[usenames]{color}
\begin{document}
\title{Multifractality can be a universal signature of phase transitions}

\author{Zhi Chen} 
\affiliation{Department of Modern physics, University of Science and
  Technology of China, Hefei, Anhui 230026, China}
\author{Xiao Xu} 
\affiliation{Department of Modern physics, University of Science and
  Technology of China, Hefei, Anhui 230026, China}



\begin{abstract}
Macroscopic systems often display phase transitions where certain
physical quantities are singular or self-similar at different
(spatial) scales. Such properties of systems are currently
characterized by some order parameters and a few critical
exponents. Nevertheless, recent studies show that the multifractality,
where a large number of exponents are needed to quantify systems,
appears in many complex systems displaying self-similarity. Here we
propose a general approach and show that the appearance of the
multifractality of an order parameter related quantity is the
signature of a physical system transiting from one phase to
another. The distribution of this quantity obtained within suitable
(time) scales satisfies a $q$-Gaussian distribution plus a possible Cauchy
distributed background. At the critical point the $q$-Gaussian shifts
between Gaussian type with narrow tails and L$\acute{\text{e}}$vy type
with fat tails. Our results suggest that the Tsallis $q$-statistics,
besides the conventional Boltzmann-Gibbs statistics, may play an
important role during phase transitions.




\end{abstract}
\maketitle 


Phase transitions are ubiquitous in physical systems. A first general
theory of phase transitions is the mean field theory, from which in
the vicinity of the critical point the free energy can be expanded in
a power series of a universal quantity called ``order
parameter''~\cite{Plischke2006}. This quantity fluctuates around zero
in a ``disordered'' phase above the critical point, and is non-zero in
an ``ordered'' phase below the critical point representing the broken
symmetry. Unfortunately the predictions from this theory are often not
correct when comparing with experiments. To remedy this, much progress
has been made in the theory of phase transitions during last fifty
years with the landmark of the discovery of some universal scaling
laws and the introduction of the renormalization group
method~\cite{Huang1987,Binney92,MEFisherRMP1998}.  With them the
seemingly complex phase transitions in a variety of systems can be
quantified by a few critical exponents which represent the
self-similar characteristics of systems during transitions. Recently,
such ideas of statistical physics have been applied to many complex
systems in various fields and have extensively expanded the
understanding to these systems~\cite{KwapienPhysRep12}. Yet, these
studies indicate that in many cases a few scaling exponents
representing the self-similarity fail to quantify certain property of
systems. Instead, one needs a large number (spectrum) of scaling
exponents to clarify the characteristic of systems which represents a
higher level of complexity. Examples include the processes of
diffusion-limited aggregation (DLA)~\cite{AmitranoPRB91}, turbulence
and chaos~\cite{JensenPRL85,MuzyPRL91}, Human heartbeat
dynamics~\cite{Plamennature1999}, climate change~\cite{AshkenazyGRL03}
and many financial quantities such as stock
prices~\cite{KwapienPhysRep12}. This characteristic has been called
``multifractality''.  A natural question is thus whether such
behaviour can be seen in phase transitions. During last two decades
people have verified the multifractality of certain quantities on some
specific transitions such as random field transitions~\cite{BenePRA88}
and Anderson transitions at the critical
point~\cite{CastellaniJPhysA86,MirlinPRL06,RodriguezPRL09,RodriguezPRL10}.
However, a general interpretation is still absent on how the
multifractality appears with phase transitions. Especially, is the
multifractality a universal property of phase transitions as those
scaling laws?  To address these questions, we argue that one should
investigate the property of systems which appears universally in phase
transitions. The only ideal quantity to our knowledge is the order
parameter.

Another important progress recently in statistical physics is
non-extensive Tsallis $q$-statistics~\cite{TsallisJSP88,TsallisPRL95},
which is a generalization of the conventional Boltzmann-Gibbs
statistics. Specifically, when the effective number of degrees of
freedom is small, one obtains the Tsallis $q$-statistics following the
same arguments to obtain the Boltzmann-Gibbs
statistics~\cite{BarangerPhysicaA02,HanelEPL11}. In this frame some variables
such as the entropy and the energy, may be non-extensive when systems
have long-range correlations~\cite{TsallisBJP99}. It is well-known
that the correlation length is infinite at the critical point of a
continuous transition. Thus, systems near the critical point are
ideal candidates to observe such non-extensive behaviour. Just as the
limit distribution of the sum of independent and identically
distributed variables is Gaussian or L$\acute{\text{e}}$vy-stable
distribution, in $q$-statistics where the long-range correlation is
present we observe a $q$-Gaussian distribution~\cite{UmarovMJM08}.
The distributions related to Tsallis $q$-statistics have been observed
in many experiments, such as cold atoms in dissipative optical
lattices~\cite{DouglasPRL06}, superdiffusion~\cite{LiuPRL08}, solar
plasma dynamics~\cite{Burlaga09}, spin glass
relaxation~\cite{PickupPRL09}, tissue radiation
response~\cite{Sotolongo-GrauPRL10}, and financial
signals~\cite{BorlandPRL02,KwapienPhysRep12}. 

Here we show that, at temperatures near the critical point the
distribution of an order parameter related quantity satisfies a
$q$-Gaussian distribution plus a possible Cauchy background. At the
critical point, the distribution shifts between the
L$\acute{\text{e}}$vy regime and the Gaussian
regime~\cite{NakaoPhysLettA00,DrozdzEPL09}, triggering the
multifractal behaviour and signalling the phase transition.  \\ \\
\medskip
{\bf Distribution and the measuring quantity} 
 
\noindent The $q$-Gaussian distribution of interest is a symmetric distribution
and has the following form:
\begin{eqnarray}
f_q(D) = P_0\cdot[1+(q-1)(D-D_0)^2/2{\sigma^2}]^{1\over 1-q},
\label{pow_6}
\end{eqnarray}
where $\sigma$ is a scale parameter related to the variance of the
distribution, $D_0$ indicates the peak position, and $P_0$ is a
normalization parameter. Letting $q\rightarrow 1$ one recovers the
Gaussian distribution. When $q \in [1, 5/3)$, the distribution is in
the Gaussian regime. The signal is monofractal and can be described
by a single exponent.
The distribution falls into the L$\acute{\text{e}}$vy regime with
infinite variance when $q\in ({5/3},3)$, and the critical value is
$q_c={5/3}$~\cite{TsallisPRL95}. In this regime $f_q(D)$ decays
asymptotically as that of a L$\acute{\text{e}}$vy-stable
distribution. Specifically the Cauchy distribution corresponds to
$f_2(D)=\gamma/\{\pi [(D-D_0)^2+\gamma^2]\}$ where
$\gamma=\sqrt{2}\sigma$ is a scale parameter. It has been shown that
the signal with the $q$-Gaussian distribution in the
L$\acute{\text{e}}$vy regime is
multifractal~\cite{NakaoPhysLettA00,KantelhardtPhysicaA02,DrozdzEPL09}. The
multifractal behaviour is manifested by two attractors at
$(\alpha,f(\alpha))= (0,0)$ and $((q-1)/(3-q),1)$, where $\alpha$ is
the singularity strength and $f(\alpha)$ is the singularity
spectrum~\cite{DrozdzEPL09}. Further, the multifractality may survive
even when the time correlation is destroyed by shuffling the
data~\cite{DrozdzEPL09}. 

We hypothesize a universal measuring quantity which possibly satisfies
a $q$-Gaussian distribution as follows. It contains only the order
parameter and does not depend on details of a specific system,
thus should be a dimensionless quantity. We further utilize the
``universal'' self-similar property of certain physical quantities of
the system near the phase transition. This implies that such
property does not change when the measuring scale alters. Combining
these ideas we construct the desired quantity as the ratio of two
order parameters measured at different spatial scales. Its
distribution $P(D)= P(m_{1}/m_{b})$ where $m_1$ and $m_b$ are order
parameters $m$ measured at scale $1$ and scale $b>1$, respectively.

When far away from the critical point the distribution of an order
parameter should be Gaussian at different spatial scales.
Its mean is zero in a disordered phase and non-zero in an ordered
phase.
Thus in the disordered phase $P(D)$ satisfies a Cauchy
distribution. In the ordered phase, we mark the mean and the
fluctuations of the order parameter at the scale ``$1$'' (scale
``$b$'') as $x_0$ and $\delta_x(t)$ [$y_0$ and $\delta_y(t)$],
respectively.
For sufficiently large systems  $x_0 \gg \delta_x(t)$ and $y_0 \gg
\delta_y(t)$. Thus the ratio $D$ is:
\begin{eqnarray}
\frac {x_0 + \delta_x (t)}{y_0 + \delta_y (t)} &= &
\frac{x_0}{y_0} + \frac{y_0 \delta_x (t)- x_0 \delta_y (t)}{y_0^2} 
\nonumber\\ 
&& + \mathcal O\left(\frac{x_0{\delta_y (t)}^2 - y_0 \delta_x (t)\delta_y(t)}
{y_0^3}\right).
\label{pow_8}
\end{eqnarray}
From equation (\ref{pow_8}) we find that in the ordered phase $P(D)$
to the first order of $\delta$ is a Gaussian. Therefore, in two
extreme cases, $P(D)$ satisfies a $q$-Gaussian distribution. We next
investigate how $P(D)$ behaves when the system is near the critical
point from some tunable examples.  \\ \\
\medskip
{\bf Example tunable models}

\noindent We now focus on some tunable examples of finite size spin
systems to verify our hypothesis. For all models we obtain the data
through Monte Carlo simulations and apply periodic boundary conditions
to systems of different sizes. When the size $L \rightarrow \infty$
one obtains the property of a macroscopic system. The system at the
scale ``1'' is the original spin system, and the block spin system at
the scale ``$b$'' is constructed using the standard coarse-graining
operation in statistical
physics~\cite{Huang1987,Binney92,MEFisherRMP1998} (see the
  Appendix~\ref{appendix1}).

We investigate two well-known models: (1) The 2-dimensional
nearest-neighbour Ising model with the Hamiltonian $H=-J \sum_{\langle
  i j\rangle} s_i s_j$, where $\langle i j \rangle$ represents nearest
neighbours, ferromagnetic parameter $J=1$ and the spin
$s_i=\pm1$. This model has a second order phase transition at
$T_{\text{c}}=2J/[\ln(1+\sqrt{2})]$. The temperature is in units of
$1/k_{\text{B}}$ where $k_{\text{B}}$ is the Boltzmann constant. Its
order parameter is the magnetization (per site) and we obtain the time
series using the Wolff algorithm~\cite{WolffPRL89}. The ratio $D(t) =
{m(1, t)/ m(b, t)}$ where $m(b, t)$ and $m(1, t)$ are order parameters
measured at the same Monte Carlo step (MCS) ``$t$''.  (2) The
3-dimensional nearest-neighbour Ising glass model which has a second
order phase transition with very slow dynamics near the critical
point~\cite{BhattPRB88,ChenPRL10}. Its $T_{\text{c}}=0.95$ and the
Hamiltonian $H=- \sum_{\langle i j\rangle} J_{ij}s_i s_j$ where we set
$J_{ij}$ satisfying a Gaussian distribution with zero mean and unit
variance, and the spin $s_i=\pm1$. Each configuration of $J_{ij}$ is
one sample. At each temperature after the equilibrium we simulate 150
samples with at least two million MCS per sample.
The order parameter is the spin overlap $m = \left(\sum_i s^{(1)}_i
s^{(2)}_i\right)/N$ where $\{s^{(2)}_i\}$ is a replica of
$\{s^{(1)}_i\}$, $N$ is the number of spins. The order parameter time
series are obtained using the Metropolis algorithm~\cite{Landau2005}.
\begin{figure}
\epsfysize=0.99\columnwidth{\rotatebox{-90}{\epsfbox{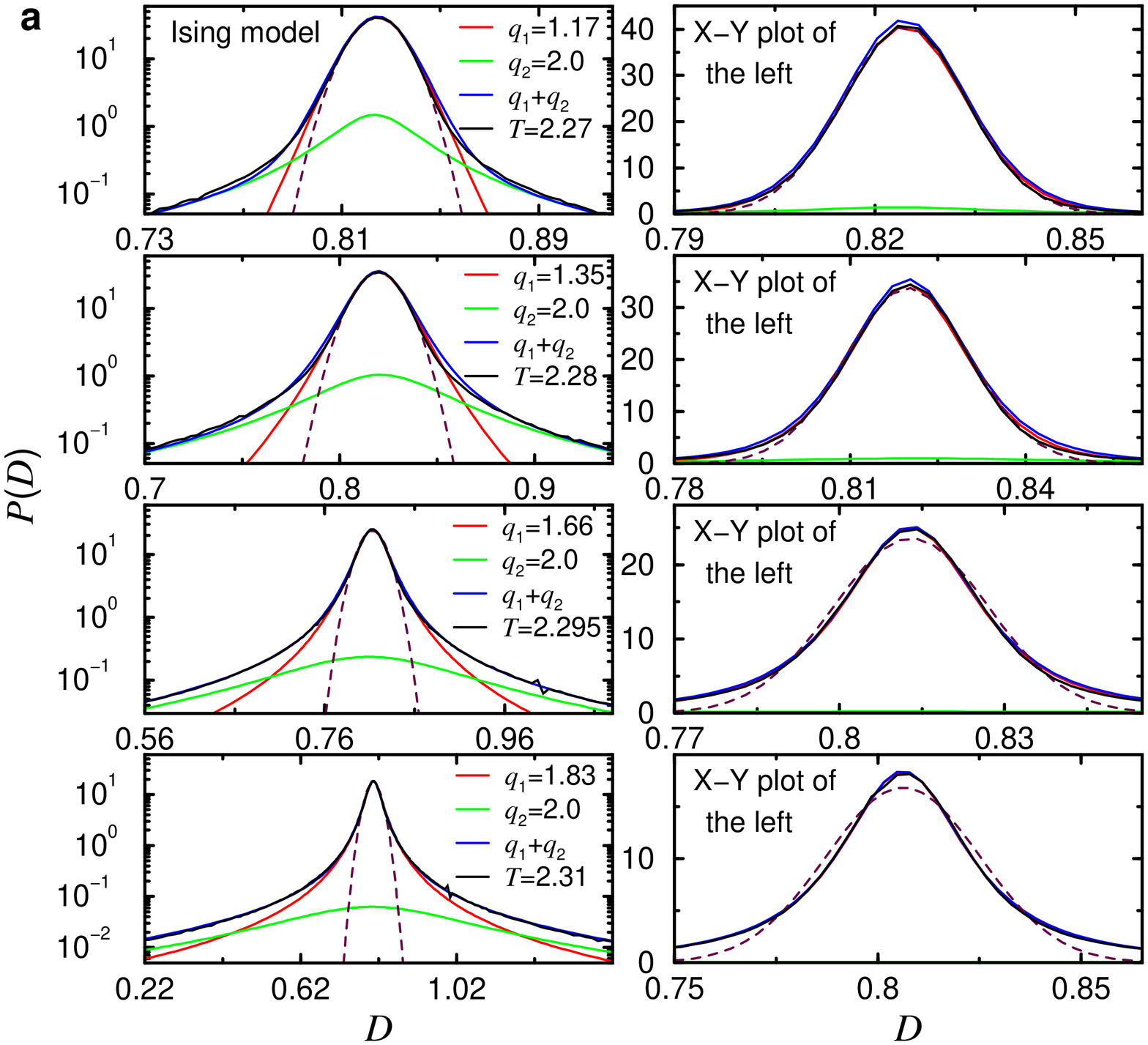}}}
\epsfysize=0.99\columnwidth{\rotatebox{-90}{\epsfbox{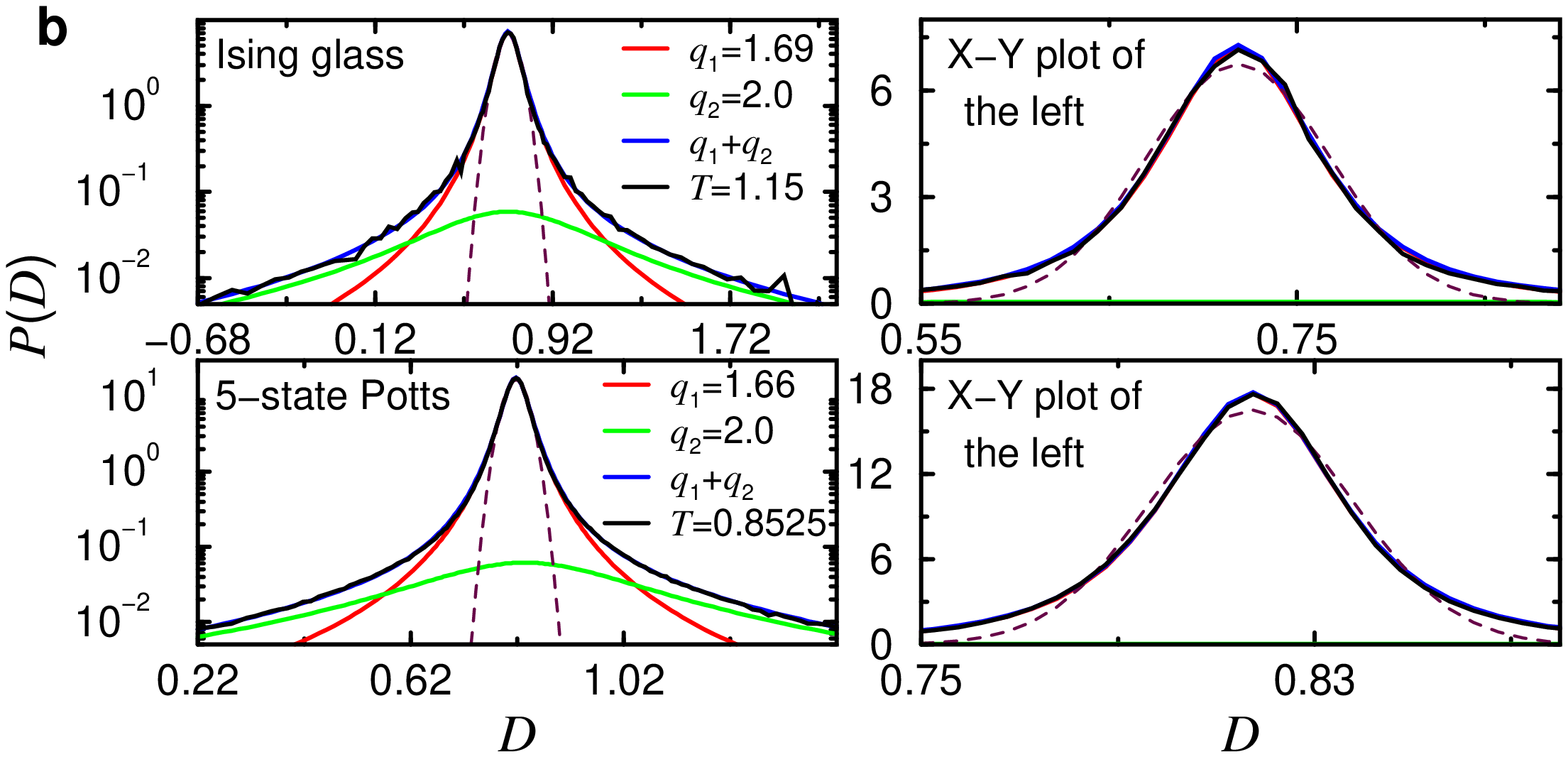}}}
\caption{Distributions of the order parameter ratio $P(D)$. {\bf a},
  At different temperatures for the 2-dimensional Ising model with the
  size $L=124$. {\bf b}, At the critical temperature $T_{\text{c},q}^L$ for
  the 3-dimensional Ising glass model with the size $L=12$, and the
  2-dimensional 5-state Potts model with the size $L=48$ and the
  thermal averaging interval $N_{\text{tw}}=80,000$. The lengths of
  the order parameter ratio for three models are 8-16 million, 150
  samples each with 2 million points, and 5.8 million,
  respectively. At all temperatures the central part of $P(D)$ can be
  well fitted by a $q$-Gaussian distribution, and the fat tails can be
  well fitted by a Cauchy distribution. Summation of the two provides
  a good fit at all ranges and at all temperatures. A Gaussian fit
  (dashed line) to the central part does not work when $T$
  approaches $T_{\text{c},q}^L$ and when $T>T_{\text{c},q}^L$. More details of the tails for
  the same distributions see the Appendix~\ref{appendix3}. }
\label{fin_fit}
\end{figure}

We also provide another model as a representing scenario for general
cases of phase transitions. The example is the 2-dimensional
nearest-neighbour $n$-state Potts model where $n$ is a positive
integer and each spin takes values $0,1,\ldots,n-1$. This model has a
second order transition for $n\le 4$ and a first order transition for
$n>4$~\cite{WuRevModPhys82}. Specifically, when $n=2$ it returns to
the Ising model. Its $T_{\text{c}}=J/[\ln(1+\sqrt{n})]$, the
Hamiltonian $H=-J \sum_{\langle i j\rangle} \delta_{s_i,s_j}$ and we
set $J=1$.  The order parameter is $m_{i,\alpha}={(n\times \langle
  \delta(s_i,\alpha) \rangle_T -1)}/{ (n -1)}$ where $i$ is the index
of a spin, $\alpha$ is the value that the spin can take and $\langle
\ldots \rangle_T$ indicates a thermal
average~\cite{WuJStatPhys88}. The order parameter series are obtained
using the Metropolis algorithm~\cite{Landau2005}. Due to the spin
symmetry, we obtain $P(D)$ utilizing all
$D_{ij}(t)=m_{i,\alpha}(1,t)/m_{j,\alpha}(b,t)$ where $j$ is the index
of the block spin, $i$ is the index of any original spin within the
block spin $j$. We have fixed the value of $\alpha$ in the
calculation.

In the simulations we set the rescaling factor $b=4$ for the Ising model
and the Potts model, and $b=2$ for the glass model.
\\
\\
\medskip
{\bf $P(D)$ near the critical point}

\noindent Results of $P(D)$ for the Ising model near the critical
point are shown in Fig.~\ref{fin_fit}a and in the
Appendix~\ref{appendix3}. When approaching the critical point from the
ordered phase ($T<T_{\text{c}}$), a Gaussian fit to $P(D)$ gradually
does not work well, and it is worse when $T>T_{\text{c}}$. In these
situations the testing Gaussian fit underestimates $P(D)$ at the peak,
overestimates $P(D)$ on two shoulders, and diminishes faster than the
exponential decay compared to the decay of $P(D)$ which contains
power-law tails. Instead, a $q$-Gaussian fit coincides very well with
the central part of $P(D)$ at all temperatures. We observe an
increasing value of the parameter $q$ with increasing temperature. At
$q=5/3$ the $q$-Gaussian enters L$\acute{\text{e}}$vy regime with
infinite variance, and we take this point as our critical temperature
$T_{\text{c},q}^L$ for the finite system with the size $L$. Thus from
this specific example, we find that near the critical point the order
parameter ratio $D$ is our desired quantity.

When the ratio $D$ is very far away from the peak $D_0$, we find that
$P(D)$ separates from such $q$-Gaussian fit.
At all temperatures we observe that fat tails of $P(D)$ decay as
$D^{-2}$, suggesting possible ``universal'' origin. Since in the
disordered phase $P(D)$ follows a Cauchy distribution, we fit these
fat tails with this distribution and take them as background. The
Cauchy fits work well even when $P(D) \sim 10^{-5}$ (see the
Appendix~\ref{appendix3}). As shown in
Fig.~\ref{fin_fit}a, the summation of a $q$-Gaussian distribution and
a suitable Cauchy background at each temperature of interest provides
good fit to $P(D)$ in all ranges of $D$, i.e., $P(D) = (1-p)\,f_q (D)
+ p \,f_{2}(D)$ where $f_{2}(D)$ is due to the Cauchy tails and $p$ is
the probability of the Cauchy contribution. This implies that the
central part and tails part of $P(D)$ may be independent. By reversing
our procedure we can deduce the origin of the Cauchy part (see the
Appendix~\ref{appendix4}). We find that the
distribution of $m(1,t)$ or $m(b,t)$ which contributes to the Cauchy
part outside the cross points with the central part of $P(D)$ achieves
a local maximum and is symmetric about zero, indicating a
disorder-like behaviour. It decays with approximately exponential
tails for large values of $m$ and decays faster for the larger system.

Similar considerations can be done on both the Ising glass model and
the Potts model. We find that the ratio distributions of them both
share the similar behaviour with that of the Ising model. The Gaussian
distribution could not fit the central part of $P(D)$ at temperatures
near the critical temperature $T_{\text{c},q}^L$. This is true even for the
5-state Potts model which has a first order phase transition.  In
Fig.~\ref{fin_fit}b we show the fits at $T_{\text{c},q}^L$ for both
models. When approaching $T_{\text{c},q}^L$ we observe Cauchy distributed fat
tails where the ratio $D$ is far away from the peak $D_0$. Further,
combining a $q$-Gaussian distribution and a suitable Cauchy background
provides a good fit to $P(D)$ at all ranges of $D$. Nevertheless, for
the Potts model, one has to choose a suitable thermal averaging
interval $N_{\text{tw}}$ (in units of MCS) for the order parameter. For
the 5-state Potts model with the size $L=48$ we take
$N_{\text{tw}}=80,000$ at all temperatures. 
\\
\\
\medskip
{\bf General scenario}

\begin{figure}
\epsfysize=0.99\columnwidth{\rotatebox{-90}{\epsfbox{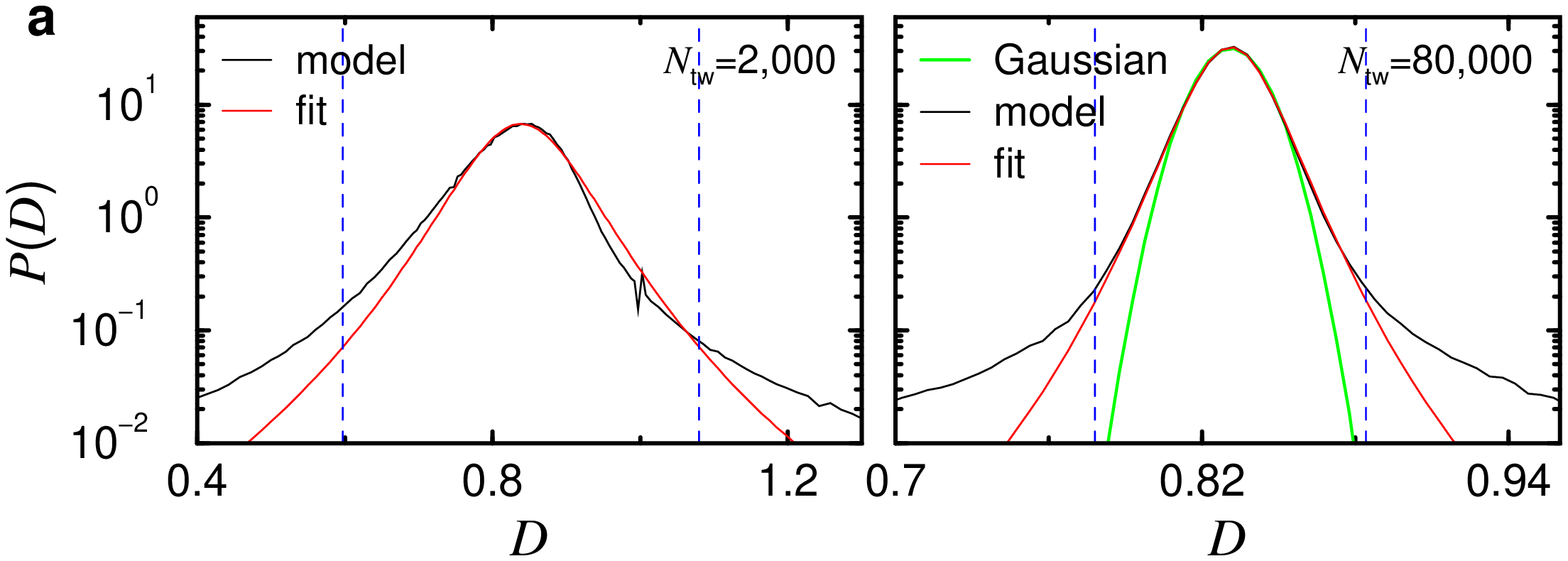}}}
\epsfysize=0.99\columnwidth{\rotatebox{-90}{\epsfbox{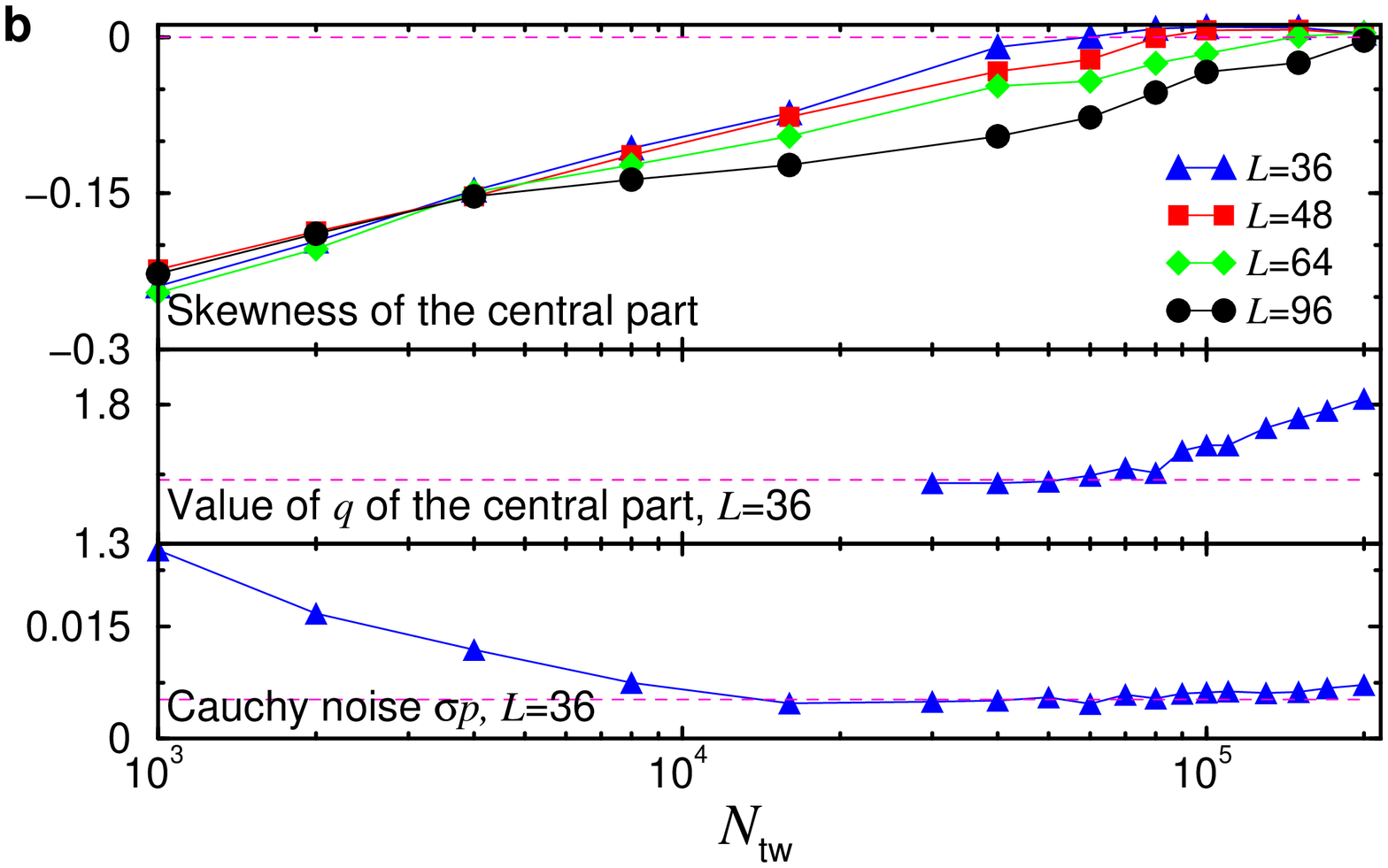}}}
\caption{Characteristics of $P(D)$ for the 5-state Potts model
  measured with different thermal averaging interval
  $N_{\text{tw}}$. {\bf a}, Asymmetric and symmetric distributions
  with different $N_{\text{tw}}$ for the 5-state Potts model at
  $T=0.8495$ ($q=1.3$). The system size is $L=48$ and the red lines
  are $q$-Gaussian fits. {\bf b}, Dependence of different parameters
  on the value of $N_{\text{tw}}$ where we fix the temperature
  $T=0.8515$.  }
\label{Cauchy_contribute}
\end{figure}

\noindent As shown in Fig.~\ref{Cauchy_contribute}a, $P(D)$ taken with
randomly chosen $N_{\text{tw}}$, e.g., $N_{\text{tw}}=2,000$ for the
5-state Potts system with the size $L=48$, may be asymmetric thus
would not follow a $q$-Gaussian distribution. 
With the Potts model as an example, here we propose an approach which
can be applied to general physical systems.

\begin{figure}
\epsfysize=0.99\columnwidth{\rotatebox{-90}{\epsfbox{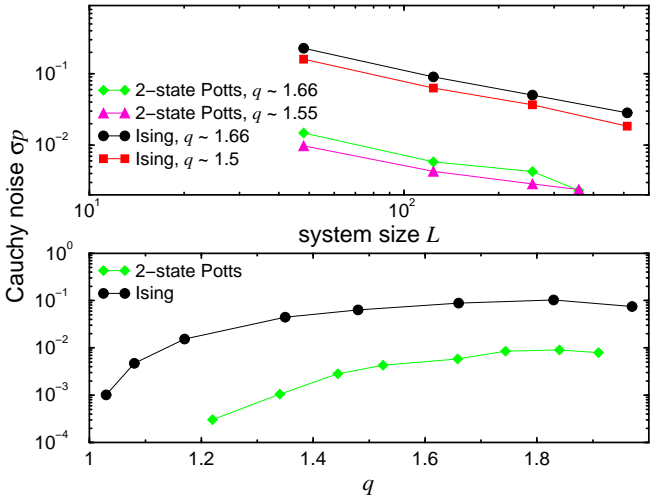}}}
\caption{Characteristics of the Cauchy noise strength in $P(D)$ within
  suitable range of $N_{\text{tw}}$. The Cauchy noise strength is
  proportional to the $\sigma p$ of the noise where $D$ is far away
  from the peak $D_0$ and is invariant within suitable range of
  $N_{\text{tw}}$. Here we show the dependence of the $\sigma p$ of
  the Cauchy noise on the system size $L$ for fixed $q$, and its
  dependence on the value of $q$ for fixed $L=124$ and
  $N_{\text{tw}}=15,000$.
}
\label{Cauchy_contribute2}
\end{figure}

First, from the construction of our measuring quantity we hypothesize
that the possible ``universal'' $q$-Gaussian distribution is relevant
to the self-similarity of certain physical quantities near the
transition. To verify this, we investigate the self-similarity of the
distribution $P(m)$ of the order parameter $m$ at different spatial
scales and find that it is broken for small values of $N_{\text{tw}}$
(see the Appendix~\ref{appendix1}). Correspondingly we obtain asymmetric
$P(D)$. The self-similarity of $P(m)$ becomes better for larger values
of $N_{\text{tw}}$. When $N_{\text{tw}}=150,000$ for the 5-state Potts
model with the size $L=96$ the distributions $P(m)$ at two scales are
almost identical after the rescaling. Correspondingly $P(D)$ becomes
symmetric.

We thus next quantify how $P(D)$ responds when the value of
$N_{\text{tw}}$ alters. To do this, we calculate the skewness of the
central part of $P(D)$. As examples shown in Fig.~\ref{Cauchy_contribute}a, 
the considered region is within two dashed lines. 
We fix the temperature and study how the skewness changes with
different values of $N_{\text{tw}}$ (see the Appendix~\ref{appendix5}
for relevant details). For different sizes of systems we find that the
skewness is closer to zero for larger value of $N_{\text{tw}}$. For
sufficiently large $N_{\text{tw}} \geq N_{\text{tw}}^{\text{min}}$ the
skewness stays at around zero, while $P(D)$ is symmetric and its
central part follows a $q$-Gaussian distribution. The value of
$N_{\text{tw}}^{\text{min}}$ seems larger for larger size of systems
and it is comparable with the characteristic time scale of the order
parameter $m$ (see the Appendix~\ref{appendix6}). Specifically, for
the 2-state Potts model $N_{\text{tw}}^{\text{min}}$ is around the
number of spins (see the Appendix~\ref{appendix5}), thus the results
of the 2-state Potts model is consistent with that of the Ising
model. Starting from $N_{\text{tw}}^{\text{min}}$, some specific
short-range properties of the Potts model are lost and the $q$ value
from a $q$-Gaussian fit to $P(D)$ keeps constant in a broad range of
$N_{\text{tw}}$ (see Fig.~\ref{Cauchy_contribute}b).  Thus we consider
the value of $q$ obtained with $N_{\text{tw}}$ chosen in such range as
the suitable value of $q$. When $N_{\text{tw}}$ further increases,
however the value of $q$ will increase. In this situation the order
parameter itself is gradually not a good quantity to quantify the
system. The above are remarks how we should choose the order
parameters.

Finally we evaluate the effect of the Cauchy noise which appears for
all models. It is pronounced where $D$ is far away from $D_0$ and its
contribution is $pf_2(D) \sim \sigma p /(D-D_0)^2$. We thus consider
$\sigma p$ as a reliable quantity to measure the strength of the
Cauchy noise. In the suitable range of $N_{\text{tw}}$, the $\sigma p$
of the Cauchy noise is invariant (see Fig.~\ref{Cauchy_contribute}b
and the Appendix~\ref{appendix5}). We then study how
it depends on the system size $L$ for fixed $q$ and suitable
$N_{\text{tw}}$. As examples shown in Fig.~\ref{Cauchy_contribute2},
for both the 2-state Potts model from our general approach and the
Ising model we find that it decays in power-law with the system
size. This implies that for a macroscopic system the Cauchy
contribution may go to zero and $P(D)$ is $q$-Gaussian in all ranges
of $D$. We also study how the $\sigma p$ of the Cauchy noise depends
on the parameter $q$ for fixed $L$ and $N_{\text{tw}}$. By comparing
the results of two equivalent models mentioned above, we find that
$P(D)$ from the former contains weaker Cauchy noise. Further, we also
obtain better $q$-Gaussian fit to $P(D)$ of the Potts model which is
constructed with fewer points (see Fig.~\ref{fin_fit}).
\begin{figure}
\epsfysize=0.99\columnwidth{\rotatebox{-90}{\epsfbox{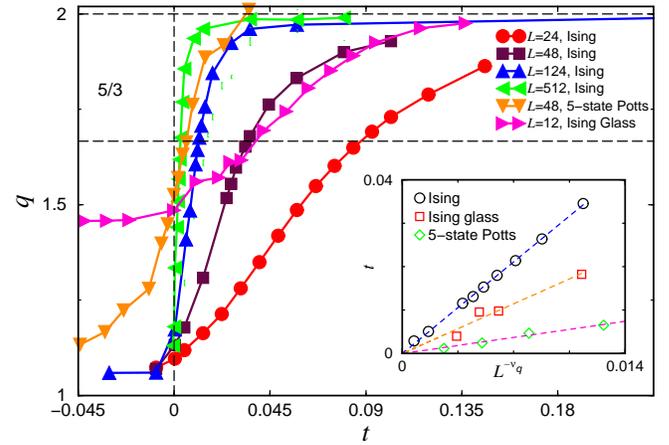}}}
\caption{ Critical behaviour for all three models. The systems enter
  the L$\acute{\text{e}}$vy regime when $q\ge q_c = 5/3$. When
  $q<5/3$, the systems are in the Gaussian regime. For better view we
  have made some scale transformations: $t\rightarrow 5t$ for the
  5-state Potts (with $N_{\text{tw}}=80,000$) and $t\rightarrow t/5$
  for the glass. Inset: Finite size scaling for all three models. The
  dashed lines are fitting lines. We show how the size-dependent
  critical temperature $T_{\text{c},q}^L$ depends on the system size $L$.  We
  mark $t=(T_{\text{c},q}^L-T_{\text{c},q})/T_{\text{c},q}$ where $T_{\text{c},q}=T_{\text{c}}$. We have
  made some scale transformations: $t \rightarrow 4t$ and $L^{-\nu_q}
  \rightarrow 4L^{-\nu_q}$ for the 5-state Potts model; $t \rightarrow
  t/20$ and $L^{-\nu_q} \rightarrow L^{-\nu_q}/10$ for the Ising glass
  model. We observe power-law dependence for all models.}
\label{q_t}
\end{figure}
\medskip
\\
{\bf Critical behaviour and multifractality}
\medskip

\noindent We now subtract the Cauchy noise and investigate the rest of
$P(D)$, whose behaviour is related to macroscopic systems.  In
Fig.~\ref{q_t} we show for all three models how the value of $q$ from
a $q$-Gaussian fit depends on the reduced temperature $t=(T-T_c)/T_{\text{c}}$,
where $T_{\text{c}}$ is obtained from the conventional methods. For each finite
system with the size $L$, we search for its corresponding critical
temperature $T_{\text{c},q}^L$. We employ this on different sizes of systems
and all models. From the finite size scaling we have $T_{\text{c},q}^L =
a{L^{-\nu_q}}+T_{\text{c},q}$.  We take the values of $T_{\text{c},q}$ as those from
conventional methods and find good power-law behaviours between
$T_{\text{c},q}^L-T_{\text{c},q}$ and $L$ (see inset of Fig.~\ref{q_t}).  The
critical exponents $\nu_q$ for the Ising model, the Ising glass model
and the 5-state Potts model are 1.16, 1.22 and 1.61,
respectively. Thus the critical temperatures from our method is
consistent with those from conventional considerations.

\begin{figure}
\epsfysize=0.99\columnwidth{\rotatebox{-90}{\epsfbox{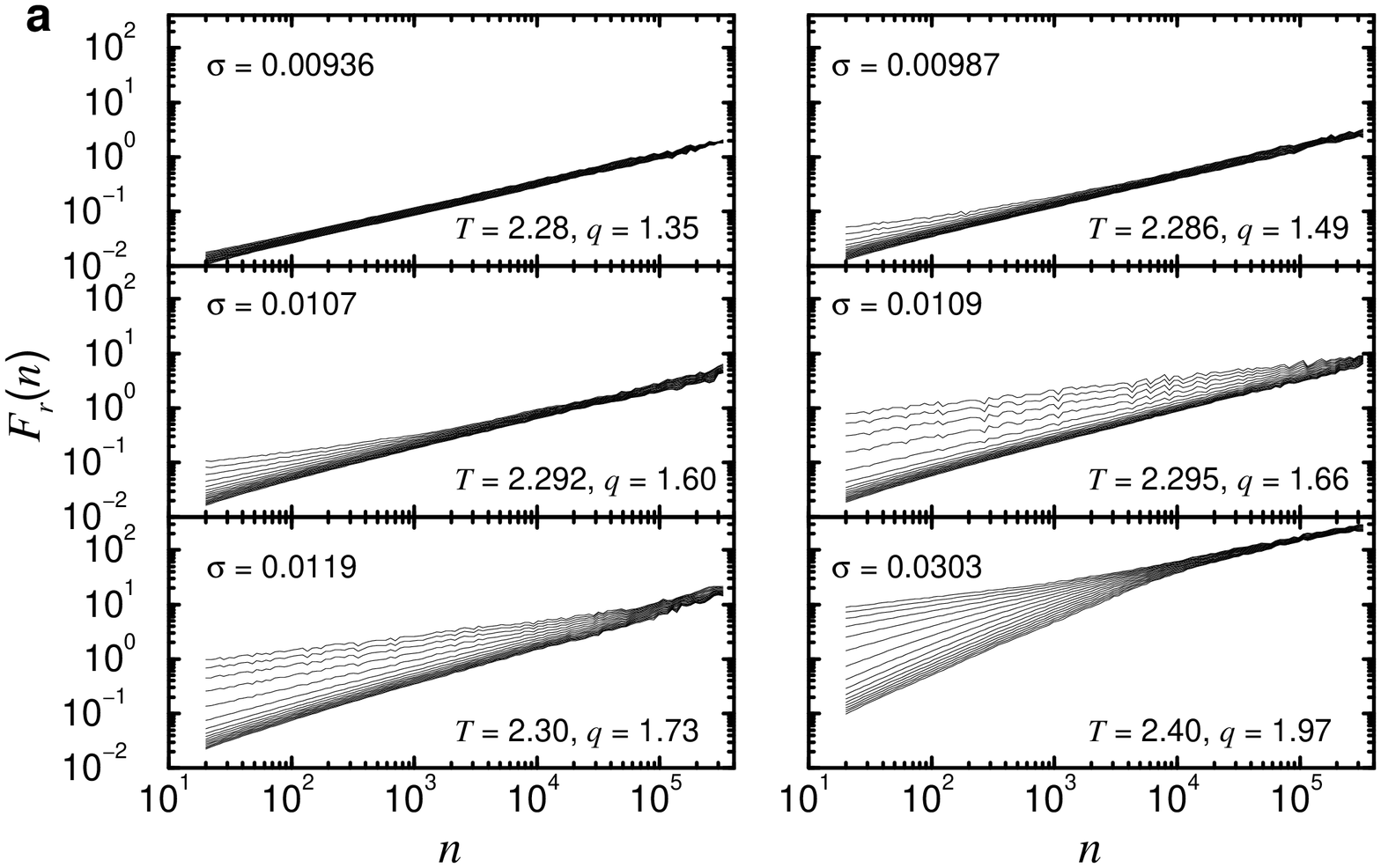}}}
\epsfysize=0.99\columnwidth{\rotatebox{-90}{\epsfbox{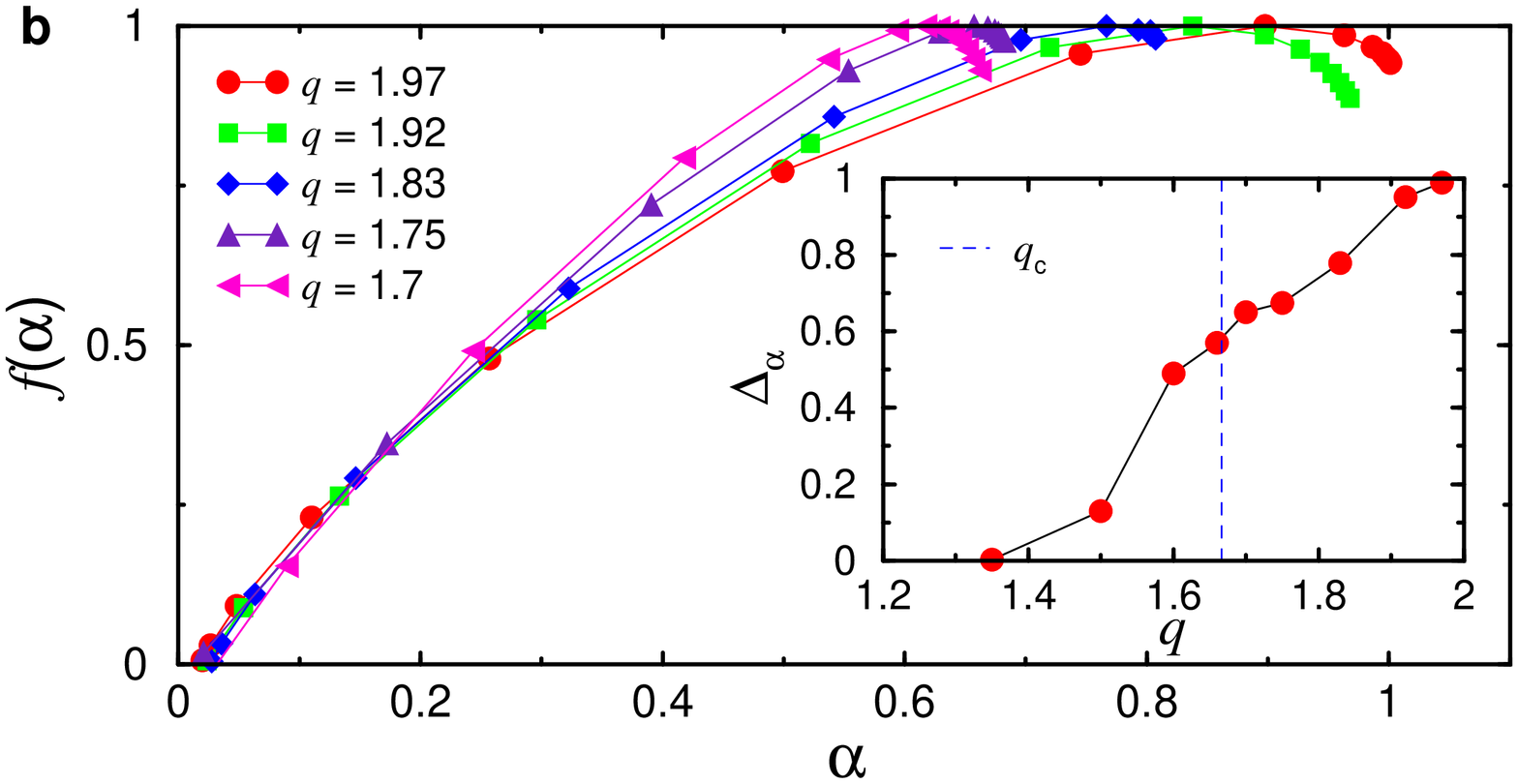}}}
\caption{ Multifractal behaviour of the Ising model near the critical
  point. The system size is $L=124$ and we have applied the MFDFA. The
  data have been shuffled before the calculation and the length of the
  order parameter ratio is 2 million. {\bf a}, The fluctuation
  function $F_r(n)$ versus observing window $n$ at different
  temperatures. Both the parameters $q$ and $\sigma$ are obtained from
  the $q$-Gaussian fit to the central part of $P(D)$. $F_r(n)$ are
  calculated in the ranges of $r \in [-4,4]$ and $n \in
  (20, 327,680)$. {\bf b}, The singularity spectrum $f(\alpha)$ versus
  the singularity strength $\alpha$ for different values of $q$. The
  fitting range of $n$ to obtain $\alpha$ and $f(\alpha)$ is
  (20, 10,240). Inset: the range of the spectrum
  $\Delta_{\alpha}=\alpha_{\text{max}}-\alpha_{\text{min}}$ for
  different values of $q$.}
\label{mf2}
\end{figure}

We next quantify the multifractality in signals using the multifractal
detrended fluctuation analysis
(MFDFA)~\cite{PengPRE94,ChenPRE02,KantelhardtPhysicaA02,ChenPRE05} (see the
  Appendix~\ref{appendix7}), which is considered to
be a reliable method to quantify the multifractal characteristic in a
nonstationary series~\cite{KwapienPhysRep12}. With this method we can
calculate the $r$-th order fluctuation function $F_r(n)\sim n^{h(r)}$,
where $n$ is the size of the observing window. Varying $h(r)$
indicates the multifractality. We also calculate widely-used the
singularity strength $\alpha$ and the singularity spectrum
$f(\alpha)$. When $(\alpha,f(\alpha))$ spreads from one point to a
variety of points in the X-Y plane, the multifractal behaviour
appears.

In Fig.~\ref{mf2} with the Ising model as an example we show the
multifractal behaviour of the shuffled series of the order parameter
ratio where the time correlation is absent. Similar behaviour is also
seen on other models (see the Appendix~\ref{appendix8}). We find that
the order parameter ratio is monofractal when the corresponding $q$ is
far below $q_c=5/3$. When we approach $T_{\text{c},q}^L$ from the
ordered phase, the slope of $F_r(n)$ gradually shifts from small
scales of $n$ for certain values of $r$. Yet at large scales all
fluctuation functions still maintain the identical slope. At $q=q_c$
the multifractality appears at all scales we observe. Nevertheless,
the range of $n$ where the multifractality presents shrinks when
continuing increasing the value of $q$. We further show the
singularity spectrum $f(\alpha)$ v.s. $\alpha$. We find that the
spreading range of the spectrum
$\Delta_{\alpha}=\alpha_{\text{max}}-\alpha_{\text{min}}$ is small for
small $q < 5/3$, indicating a Gaussian-like monofractal
behaviour. However when $q$ approaches $5/3$, $\Delta_{\alpha}$
rapidly increases and then slowly expands with increasing $q$.

The multifractal behaviour shown above should not be system specific
since we have considered the shuffled series. To further verify this,
we generate some $q$-Gaussian distributed artificial signals and set
the same parameters for the artificial signals and the testing model
(see the Appendix~\ref{appendix9}). We find that the results are
almost identical for the artificial signals and the model. Thus such
multifractal behaviour is determined by the two parameters $q$ and
$\sigma$ of the $q$-Gaussian distribution. To observe the
multifractality, it requires $q\ge 5/3$. When $q>5/3$, the range where
the multifractality presents is controlled by $\sigma$. For fixed
value of $q$, the larger size of system, the smaller value of $\sigma$
(see the Appendix~\ref{appendix9}). Correspondingly the larger system
presents multifractal behaviour in broader range of $n$.
\medskip
\\
\\
{\bf Conclusion} 
\medskip
\\
\noindent With tunable examples we have shown that physical systems
near the phase transition present much more complex self-similar
behaviour represented by the multifractality of the ratio of two order
parameters measured at different spatial scales. At all temperatures
around the critical point the distribution of the ratio follows a
non-extensive $q$-Gaussian distribution plus a possible Cauchy
background which decays in power-law with the system size and may
disappear for a macroscopic system. The $q$-Gaussian distribution
enters the L$\acute{\text{e}}$vy regime at the critical point, and
triggers the multifractality at all scales we observe. We have
proposed a general approach which relates only to the broken symmetry
yielding zero and non-zero order parameters in different phases as
well as the self-similar characteristics of systems near the
transition. Thus it should be applicable to other physical
systems. The multifractality appears for the order parameter $m$
obtained within suitable range of (time) scales where certain
short-range properties of the specific system are lost and $m$ shows
good spatial self-similarity.  In this situation the long-range
correlation in the order parameter prevails and the ratio of order
parameters follows a non-extensive q-Gaussian distribution. Our
results suggest that the Tsallis q-statistics may play an important
role in phase transitions.

This work is supported by the National Natural Science Foundation of
China (Grant no. 11275184), Anhui Provincial Natural Science
Foundation (Grant no. 1208085MA03), and the Fundamental Research Funds
for the Central Universities of China (Grants no. WK2030040012 and
2340000034).
%
%
%
%


\begin{thebibliography}{99}


\bibitem{Plischke2006} Plischke, M. \& Bergersen, B. {\it Equilibrium Statistical Physics, 3rd Edition} (World Scientific, New Jersey, 2006).

\bibitem{Huang1987} Huang, K. {\it Statistical Mechanics, 2nd Edition} (John Wiley \& Sons, New York, 1987).

\bibitem{Binney92} Binney, J.J., Dowrick, N.J., Fisher, A.J. \& Newman, M.E.J. {\it The Theory of Critical Phenomena: An Introduction to the Renormalization Group} (Oxford Univ. Press, New York, 1992).

\bibitem{MEFisherRMP1998} Fisher, M.E. Renormalization group theory: Its basis and formulation in statistical physics. {\it Rev. Mod. Phys.} {\bf 70}, 653-681 (1998).

\bibitem{KwapienPhysRep12} Kwapien, J. \& Drozdz, S. Physical approach to complex systems. {\it Phys. Rep.} {\bf 515}, 115-226 (2012).

\bibitem{AmitranoPRB91} Amitrano, C., Coniglio, A., Meakin, P. \& Zanetti, M. Multiscaling in diffusion-limited aggregation. {\it Phys. Rev. B} {\bf 44}, 4974-4977 (1991).

\bibitem{JensenPRL85} Jensen, M.H., Kadanoff, L.P., Libchaber, A., Procaccia, I. \& Stavans, J. Global universality at the onset of chaos: Results of a forced Rayleigh-Bénard experiment. {\it Phys. Rev. Lett.} {\bf 55}, 2798-2801 (1985).

\bibitem{MuzyPRL91} Muzy, J.F., Bacry, E. \& Arneodo, A. Wavelets and multifractal formalism for singular signals: Application to turbulence data. {\it Phys. Rev. Lett.} {\bf 67}, 3515-3518 (1991).

\bibitem{Plamennature1999} Ivanov, P.Ch., {\it et al.} Multifractality in human heartbeat dynamics. {\it Nature} {\bf 399}, 461-465 (1999).

\bibitem{AshkenazyGRL03} Ashkenazy, Y., Baker, D.R., Gildor, H. \& Havlin, S. Nonlinearity and multifractality of climate change in the past 420,000 years. {\it Geophys. Res. Lett.} {\bf 30}, 2146-2149 (2003).

\bibitem{BenePRA88} Bene, J. \& Szepfalusy, P. Multifractal properties in the one-dimensional random-field Ising model. {\it Phys. Rev. A} {\bf  37}, 1703-1707 (1988).

\bibitem{CastellaniJPhysA86} Castellani, C. \& Peliti, L. Multifractal wavefunction at the localisation threshold. {\it J. Phys. A: Math. Gen.} {\bf 19}, L429-L432 (1986).

\bibitem{MirlinPRL06} Mirlin, A.D., Fyodorov, Y.V., Mildenberger, A. \& Evers, F. Exact relations between multifractal exponents at the Anderson transition. {\it Phys. Rev. Lett.} {\bf 97}, 046803 (2006).

\bibitem{RodriguezPRL09} Rodriguez, A., Vasquez, L.J. \& Romer, R.A. Multifractal analysis with the probability density function at the three-dimensional Anderson transition. {\it Phys. Rev. Lett.} {\bf 102}, 106406 (2009).

\bibitem{RodriguezPRL10} Rodriguez, A., Vasquez, L.J., Slevin, K. \& Romer, R.A. Critical parameters from a generalized multifractal analysis at the Anderson transition. {\it Phys. Rev. Lett.} {\bf 105}, 046403 (2010).

\bibitem{TsallisJSP88} Tsallis, C. Possible generalization of Boltzmann–Gibbs statistics. {\it J. Stat. Phys.} {\bf 52}, 479-487 (1988).

\bibitem{TsallisPRL95} Tsallis, C., Levy, S.V.F., Souza, A.M.C. \& Maynard, R. Statistical-mechanical foundation of the ubiquity of L$\acute{\text{e}}$vy distributions in nature. {\it Phys. Rev. Lett.} {\bf 75}, 3589-3593 (1995).

\bibitem{BarangerPhysicaA02} Baranger, M. Why Tsallis statistics? {\it Physica A} {\bf 305}, 27-31 (2002).

\bibitem{HanelEPL11} Hanel, R. \& Thurner, S. When do generalized entropies apply? How phase space volume determines entropy. {\it Europhys. Lett.} {\bf 96}, 50003 (2011).

\bibitem{TsallisBJP99} Tsallis, C. Nonextensive statistics: Theoretical, experimental and computational evidences and connections. {\it Brazilian Journal of Physics} {\bf 29}, 1-35 (1999).

\bibitem{UmarovMJM08} Umarov, S., Tsallis, C. \& Steinberg, S. On a $q$-central limit theorem consistent with nonextensive statistical mechanics. {\it Milan J. Math.} {\bf 76}, 307-328 (2008).

\bibitem{DouglasPRL06} Douglas, P., Bergamini, S. \& Renzoni, F. Tunable Tsallis distributions in dissipative optical lattices. {\it Phys. Rev. Lett.} {\bf 96}, 110601 (2006).

\bibitem{LiuPRL08} Liu, B. \& Goree, J. Superdiffusion and non-Gaussian statistics in a driven-dissipative 2D dusty plasma. {\it Phys. Rev. Lett.} {\bf 100}, 055003 (2008).

\bibitem{Burlaga09} Burlaga, L.F. \& Ness, N.F. Compressible "turbulence" observed in the heliosheath by Voyager 2. {\it ApJ} {\bf 703}, 311-324 (2009).

\bibitem{PickupPRL09} Pickup, R.M., Cywinski, R., Pappas, C., Farago, B. \& Fouquet, P. Generalized spin-glass relaxation. {\it Phys. Rev. Lett.} {\bf 102}, 097202 (2009).

\bibitem{Sotolongo-GrauPRL10} Sotolongo-Grau, O., Rodriguez-Perez, D., Antoranz, J.C. \& Sotolongo-Costa, O. Tissue radiation response with maximum Tsallis entropy. {\it Phys. Rev. Lett.} {\bf 105}, 158105 (2010).

\bibitem{BorlandPRL02} Borland, L. Option pricing formulas based on a non-Gaussian stock price model. {\it Phys. Rev. Lett.} {\bf 89}, 098701 (2002).

\bibitem{NakaoPhysLettA00} Nakao, H. Multi-scaling properties of truncated Lévy flights. {\it Phys. Lett. A} {\bf 266}, 282-289 (2000).

\bibitem{DrozdzEPL09} Drozdz, S., Kwapien, J., Oswiecimka, P. \& Rak, R. Quantitative features of multifractal subtleties in time series. {\it Europhys. Lett.}{\bf 88}, 60003 (2009).

\bibitem{KantelhardtPhysicaA02} Kantelhardt, J.W., {\it et al.} Multifractal detrended fluctuation analysis of nonstationary time series. {\it Physica A} {\bf 316}, 87-114 (2002).

\bibitem{WolffPRL89} Wolff, U. Collective Monte Carlo updating for spin systems. {\it Phys. Rev. Lett.} {\bf 62}, 361-364 (1989).

\bibitem{BhattPRB88} Bhatt, R.N. \& Young, A.P. Numerical studies of Ising spin glasses in two, three, and four dimensions. {\it Phys. Rev. B} {\bf 37}, 5606-5614 (1988).

\bibitem{ChenPRL10} Chen, Z. \& Yu, C.C. Comparison of Ising spin glass noise to flux and inductance noise in SQUIDs. {\it Phys. Rev. Lett.} {\bf 104}, 247204 (2010).

\bibitem{Landau2005} Landau, D.P. \& Binder, K. {\it A Guide to Monte Carlo Simulations in Statistical Physics, 2nd Edition} (Cambridge University Press, Cambridge, UK, 2005).

\bibitem{WuRevModPhys82} Wu, F.Y. The Potts model. {\it Rev. Mod. Phys.} {\bf 54}, 235-268 (1982).

\bibitem{WuJStatPhys88} Wu, F.Y. Potts model and graph theory. {\it J. Stat. Phys.} {\bf 52}, 99-112 (1988).

\bibitem{PengPRE94} Peng, C.-K., {\it et al.} Mosaic organization of DNA nucleotides. {\it Phys. Rev. E} {\bf 49}, 1685-1689 (1994).

\bibitem{ChenPRE02} Chen, Z., Ivanov, P.Ch., Hu, K. \& Stanley, H.E. Effect of nonstationarities on detrended fluctuation analysis. {\it Phys. Rev. E} {\bf 65}, 041107 (2002).

\bibitem{ChenPRE05} Chen, Z., {\it et al.} Effect of nonlinear filters on detrended fluctuation analysis. {\it Phys. Rev. E} {\bf 71}, 011104 (2005).

\end{thebibliography}

%
%
%
%
\renewcommand{\thefigure}{S\arabic{figure}}
 \setcounter{figure}{0}
\renewcommand{\theequation}{S.\arabic{equation}}
 \setcounter{equation}{0}
\bigskip
\begin{center}
{\large Appendix}
\end{center}

\subsection{Self-similarity in distributions of the order parameter}
\label{appendix1}
\begin{figure}
\epsfysize=0.99\columnwidth{\rotatebox{-90}{\epsfbox{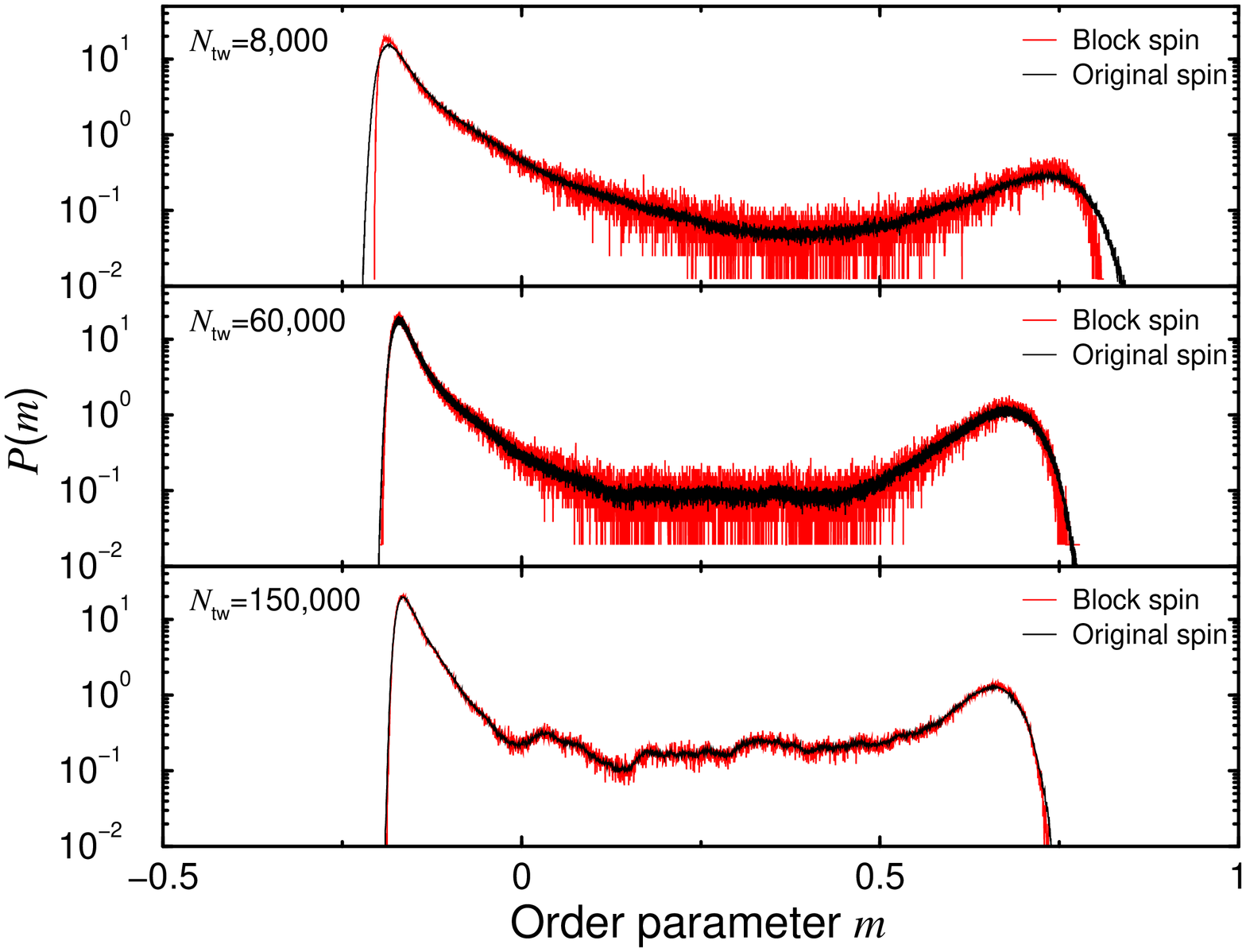}}}
\caption{Self-similarity of distributions of the order parameter
  for the 5-state Potts model. The system size is $L=96$ and the
  temperature is $T=0.8515$. For the distribution of the block
  (coarse-grained) spin system (with $L'=96/4=24$) we have done the
  following scale transformation: $x \rightarrow x/1.215, y
  \rightarrow y \times 1.215$. With larger thermal averaging interval
  $N_{\text{tw}}$ we obtain better similarity between the distribution
  of the original spin system and that of the block spin system.}
\label{appfig1}
\end{figure}
When a system is in the vicinity of the critical point, many physical
quantities of it display self-similar (scaling) properties. The
distribution of the order parameter also shares this
characteristic. In Fig.~\ref{appfig1} we show an example for the
2-dimensional 5-state Potts model with the size $L=96$. We construct
the associated block spin system with size $L'=96/4=24$ utilizing the
standard coarse-graining procedure. For a $d$-dimensional spin system
with $L^d$ lattices, we can transform it into a block spin system with
$(L/b)^d$ block spins. Each of block spins contains $b^d$ spins. A
coarse-graining operation is reached when we substitute each of block
spins by a single spin with the value determined by the majority
rule. For example, for the Ising model the block spin is +1 if there
are more spins up than down, and {\it vice versa}.  In particular,
when the amount of spins up and spins down is equal, we assign the
block spin +1 or -1 randomly. In this way we have constructed the
Ising system at the scale ``$b$''. The block spin system for the Potts
model can be constructed similarly.

Nevertheless, for the Potts model we do not see good scaling when the
thermal averaging interval $N_{\text{tw}}$ for the order parameter $m$
is too small (see Fig.~\ref{appfig1}). In this situation $m$ contains
certain short-range information specifically related to the Potts
model. It is well-known that the critical properties of a physical
system do not depend on the short-range details, but on the
characteristics of long-range fluctuations. Such short-range
information is not self-similar and may diminish with larger
$N_{\text{tw}}$ where the time correlation in the order parameter
becomes weaker. (For example, see the Appendix~\ref{appendix6}.)  For
sufficiently large value of $N_{\text{tw}}$, the distribution of the
order parameter for the original system $h_1(x)$ and that for the
block spin system $h_b(x)$ become similar, i.e.,
$h_1(x)=h_b(cx)/c$. Interestingly, when such scaling relation is
effective, the distribution of the order parameter ratio $P(D)$
becomes symmetric and follows a $q$-Gaussian distribution, as we show
in the paper.

\subsection{Cauchy noise in distributions of the order parameter ratio}
\label{appendix3}
\begin{figure}
\epsfysize=0.99\columnwidth{\rotatebox{-90}{\epsfbox{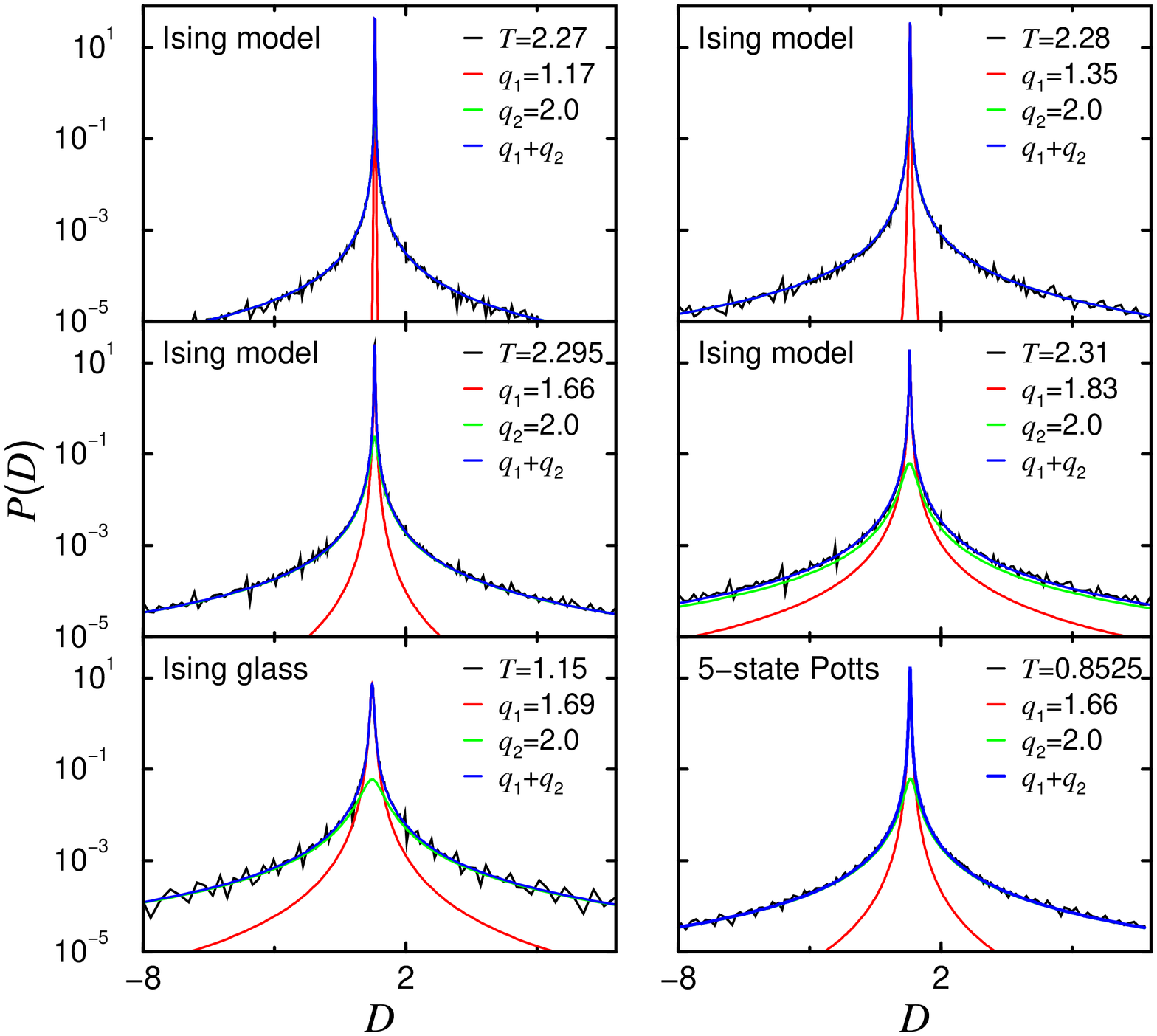}}}
\caption{Cauchy distributed fat tails in distributions of the order
  parameter ratio. We show such tails at different temperatures for
  all three models. The data are the same as those we show in Fig.~1
  of the paper.  }
\label{appfig3}
\end{figure}

In Fig.~\ref{appfig3} we present the Cauchy distributed fat tails in
distributions of the order parameter ratio $P(D)$. For the Ising model
we observe very good fit to the tails of $P(D)$ down to $10^{-4} -
10^{-5}$, while the length of the order parameter ratio is 8-16
million. We also note that at high temperatures with $q$ close to 2,
the Cauchy background may be still pronounced. In an example shown in
Fig.~\ref{appfig3} for the system at $T=2.31$ and with the size
$L=124$, we find that a single $q$-Gaussian fit works well for the
central part, however it still underestimates $P(D)$ at positions of
$D$ far away from the peak $D_0$.

The above behaviour has also been found in the Ising glass model and
the Potts model. For the former at each temperature we have 150
samples each with 2 million points after the equilibrium. For the
latter at each temperature the length of the order parameter ratio is
5.8 million. As examples in Fig.~\ref{appfig3} we show such behaviour
at the critical point $T_{\text{c},q}^L$ which we have defined in the
paper. For both models good fits to the tails of $P(D)$ down to
$10^{-4}$ are observed.

\subsection{Origin of the Cauchy noise}
\label{appendix4}
\begin{figure}
\epsfysize=0.99\columnwidth{\rotatebox{-90}{\epsfbox{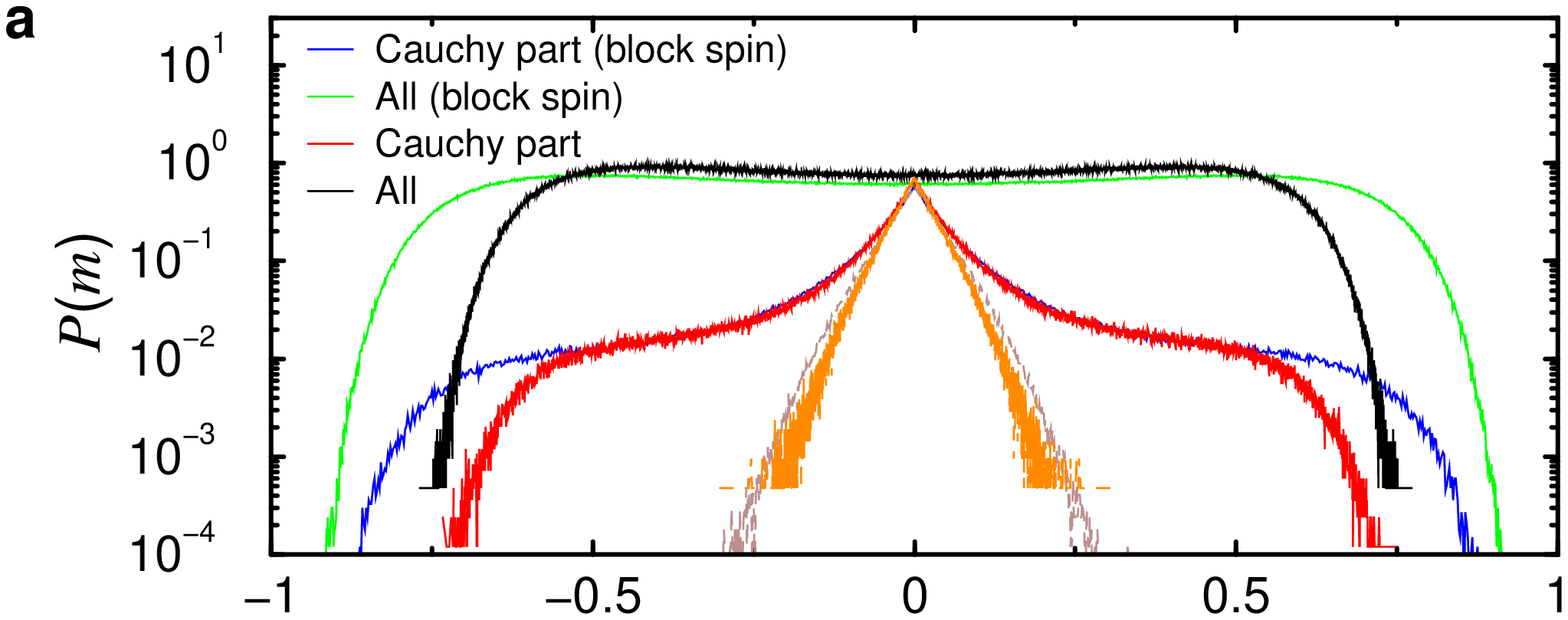}}}
\epsfysize=0.99\columnwidth{\rotatebox{-90}{\epsfbox{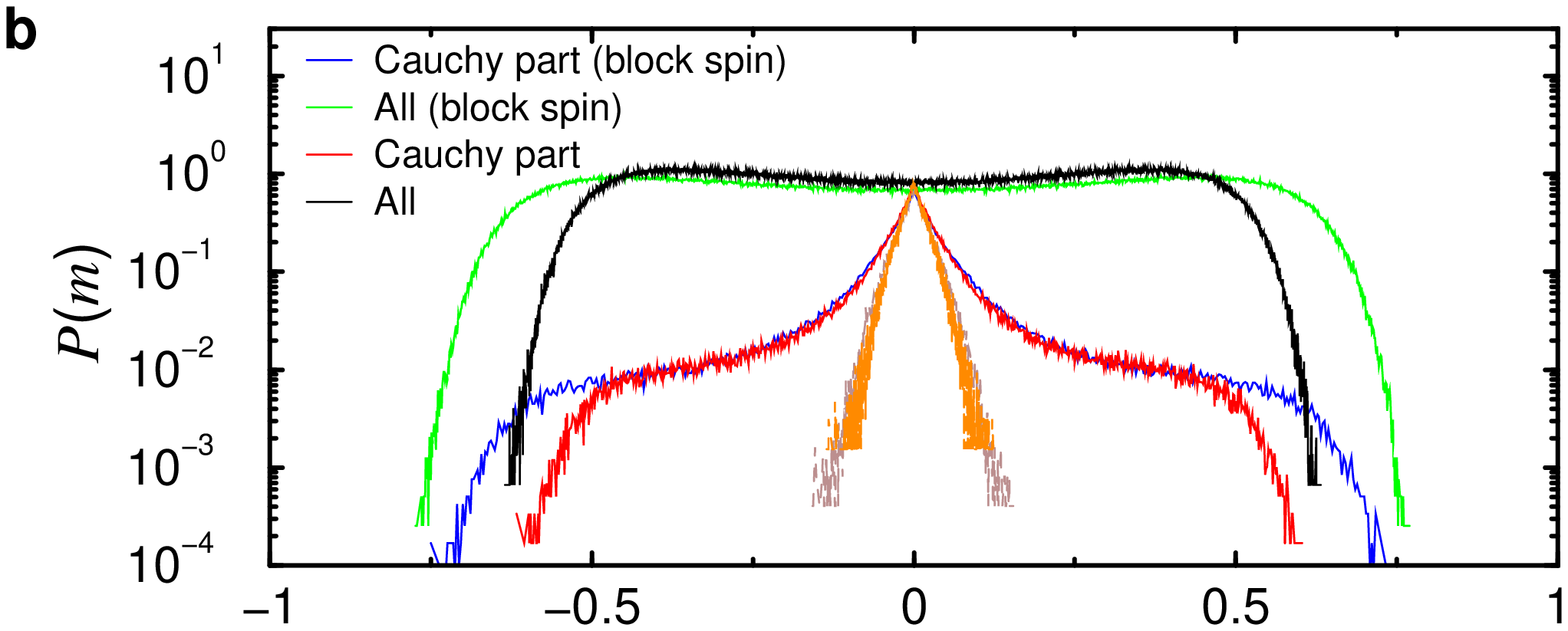}}}
\epsfysize=0.99\columnwidth{\rotatebox{-90}{\epsfbox{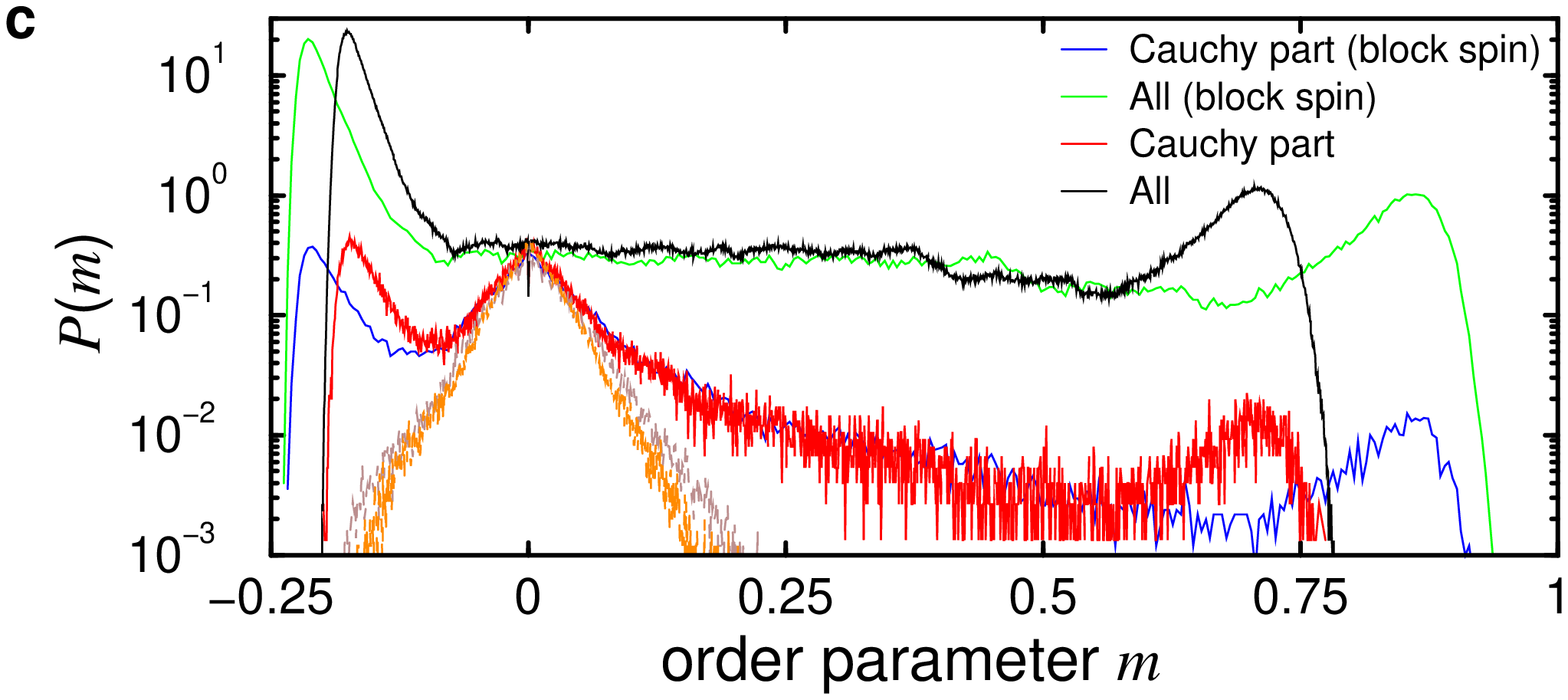}}}
\caption{Origin of the Cauchy noise as seen in the distribution $P(m)$
  of the order parameter $m$. {\bf a}, For the 2-dimensional Ising
  model at $T=2.295$ with the size $L=124$. {\bf b}, For the
  2-dimensional Ising model at $T=2.275$ with the size $L=512$. {\bf
    c}, For the 5-state Potts model at $T=0.851$ with the size $L=48$,
  while the thermal averaging interval $N_{\text{tw}}$ is 80,000. The
  values of $q$ for the distributions of the corresponding order
  parameter ratio are 1.66, 1.66 and 1.45, respectively. The orange
  and brown dashed lines are contributed by the original and block
  spins associated with the Cauchy part of $P(D)$ outside the cross
  points with the central part of $P(D)$. }
\label{appfig4}
\end{figure}

By reversing our procedure we can deduce the origin of the Cauchy
distributed fat tails. To see this, we define a quantity
$r(D)=\text{min}[f_{\text{cauchyfit}}(D)/P(D),1]$, where
$f_{\text{cauchyfit}}(D)$ is our Cauchy fit and $P(D)$ is the
distribution of the model from the Monte Carlo simulation. At each
Monte Carlo step $t$ with the order parameter ratio $D(t) = {m(1, t)/
  m(b, t)}$, we take $r(D)$ as the probability of this data point
belonging to the Cauchy noise. We did this at each $t$ and thus could
obtain the corresponding $m(1,t)$ and $m(b,t)$ which belong to the
Cauchy noise. Such considerations can be done in all ranges of $D$ or
in partial range of $D$, e.g., we can choose the range of $D$ which is
outside the cross points of the central part and the Cauchy part of
$P(D)$. As shown in the orange and brown dashed lines of
Fig.~\ref{appfig4} we find that the distribution of $m(1,t)$ or
$m(b,t)$ which is associated with the Cauchy part of $P(D)$ outside
the cross points with the central part of $P(D)$ achieves a local
maximum and is symmetric about zero, then decays with approximately
exponential tails for large values of $m$, indicating a disorder-like
behaviour. Further, the tails decay faster for the system with larger
size. This implies that the fat tails in $P(D)$ may go to zero when
the system size $L \rightarrow \infty$. Such behaviour is also true
when the distribution $P(m)$ of the order parameter is asymmetric.

\subsection{Asymmetric and symmetric $P(D)$ with different $N_{\text{tw}}$}
\label{appendix5}
\begin{figure}
\epsfysize=0.99\columnwidth{\rotatebox{-90}{\epsfbox{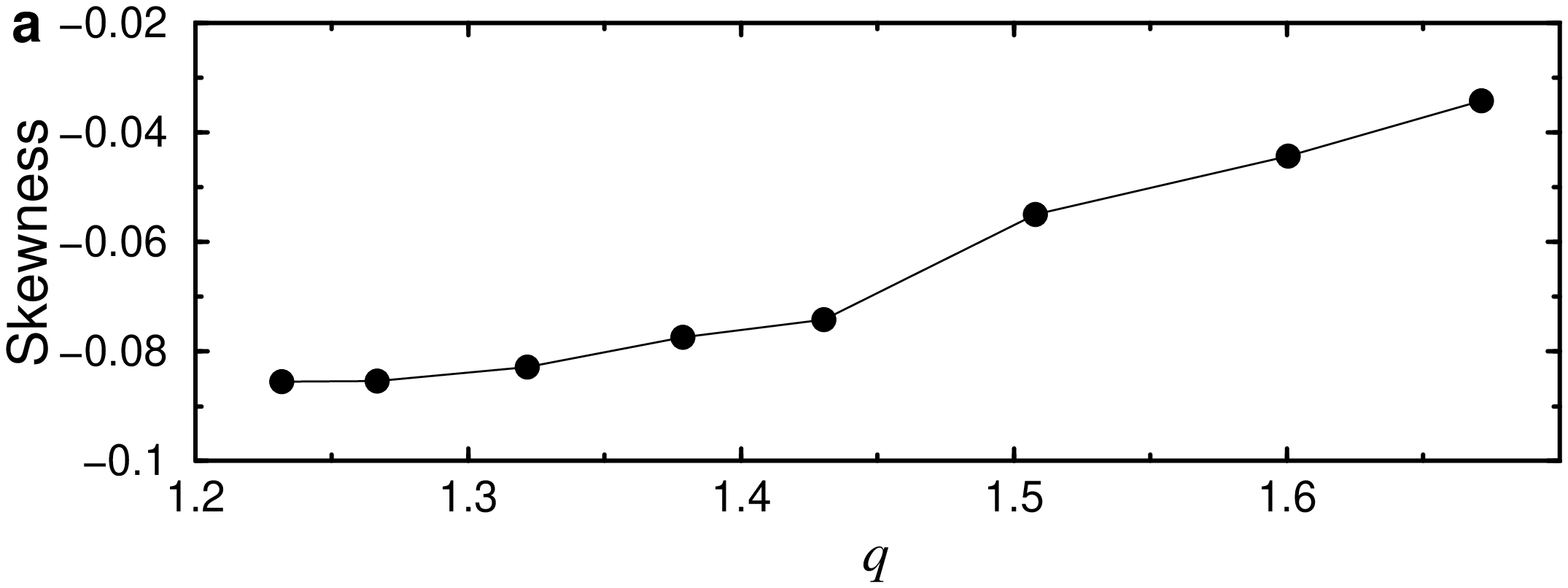}}}
\epsfysize=0.99\columnwidth{\rotatebox{-90}{\epsfbox{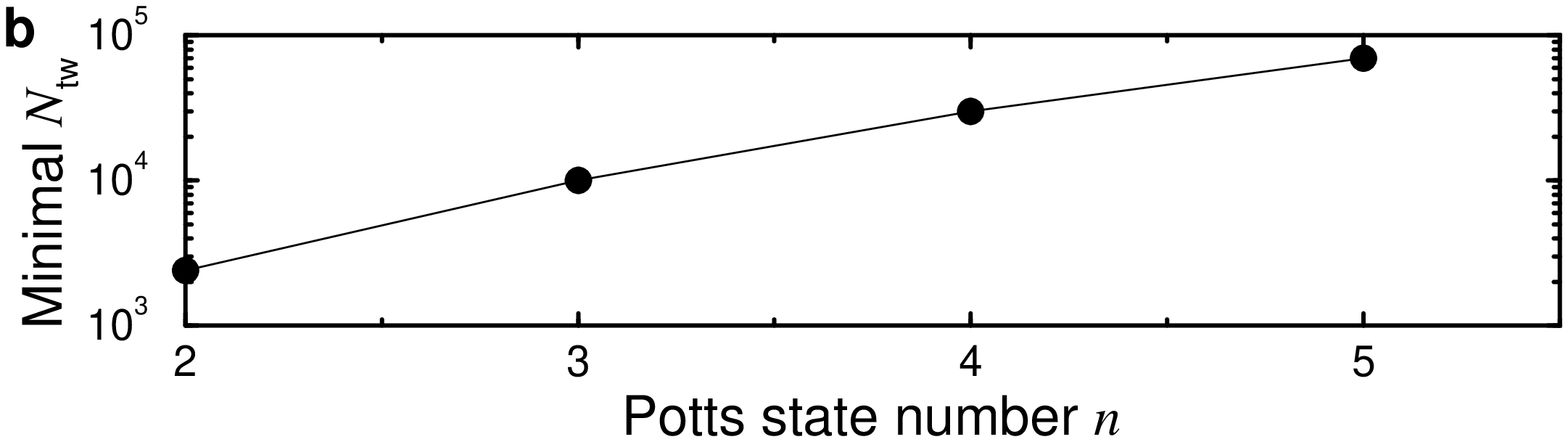}}}
\caption{Characteristics of the distributions of the order parameter
  ratio $P(D)$ for the Potts model. {\bf a}, The skewness of the
  central part of $P(D)$ for the 5-state Potts model at temperatures
  with different values of $q$. We fix the system size $L=36$ and the
  thermal averaging interval $N_{\text{tw}}=16,000$. {\bf b}, The
  dependence of the minimal $N_{\text{tw}}$ at which one could obtain
  symmetric $P(D)$ on the Potts state number $n$ at temperatures with
  fixed $q=1.5$. We fix the system size $L=48$. The skewness of the
  corresponding $P(D)$ is within (-0.01, 0.01).}
\label{appfig5}
\end{figure}
For the $n$-state Potts model with $n\geq 2$ we find that the
distribution of the order parameter ratio is not symmetric when the
thermal averaging interval $N_{\text{tw}}$ is small, as shown in
Fig.~2a of the paper. Further, such behaviour
diminishes with larger values of $N_{\text{tw}}$. To characterize
this, we can calculate the skewness of $P(D)$ around the peak
$D_0$. (In practice, the width of the region we choose to calculate is
around $11\sigma - 12 \sigma$, where $\sigma$ is the scale parameter
of the testing $q$-Gaussian fit to $P(D)$.) As shown in
Fig.~\ref{appfig5}a, such asymmetry seems larger at lower temperature
with smaller value of the parameter $q$. Further, it is also stronger
for the Potts system with more possible spin states. This is
manifested in Fig.~\ref{appfig5}b, where we fix the $q$-Gaussian fit
parameter $q=1.5$. We find that $N_{\text{tw}}^{\text{min}}$ --- the
minimal value of the thermal averaging interval $N_{\text{tw}}$ to
obtain symmetric $P(D)$ --- increases with increasing value of the
state number $n$. Specifically, we note that
$N_{\text{tw}}^{\text{min}}$ is consistent with the results of the Ising
model which is related to the 2-state case of the Potts
model. Since for $n \leq 4$ the system has a second order phase
transition, such behaviour is not related to the order of the
transition. 

\begin{figure}
\epsfysize=0.99\columnwidth{\rotatebox{-90}{\epsfbox{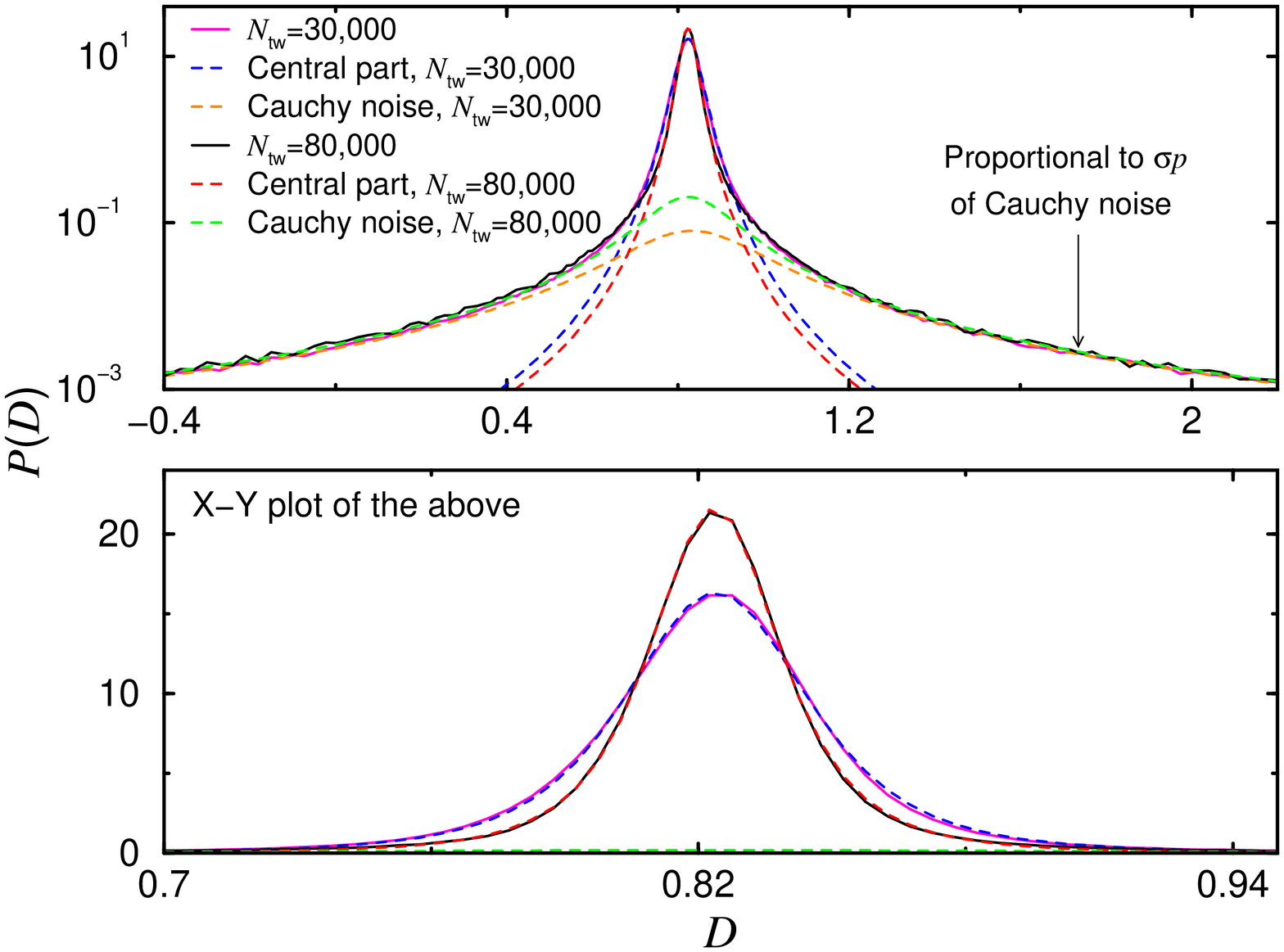}}}
\caption{Strength of the Cauchy noise for the 5-state Potts model. We
  fix the temperature $T=0.8515$ ($q\sim 1.55$) and the size
  $L=36$. For different values of $N_{\text{tw}}$ within the suitable
  range, the strength represented by the $\sigma p$ of the noise keeps
  constant.}
\label{appfig6}
\end{figure}

For sufficiently large thermal averaging interval $N_{\text{tw}}$,
$P(D)$ becomes symmetric, and it would keep symmetric when continuing
increasing the value of $N_{\text{tw}}$, as shown in
Fig.~2b of the paper. Further, the value of the
parameter $q$ in a $q$-Gaussian fit to $P(D)$ also keeps almost
invariant in a broad range of $N_{\text{tw}}$. Here we show some
examples in Fig.~\ref{appfig6}. In this suitable range the strength of
the Cauchy noise, marked by the $\sigma p$ of the noise, also keeps
constant.


\subsection{Time correlation in the order parameter time series}
\label{appendix6}
\begin{figure}
\epsfysize=0.99\columnwidth{\rotatebox{-90}{\epsfbox{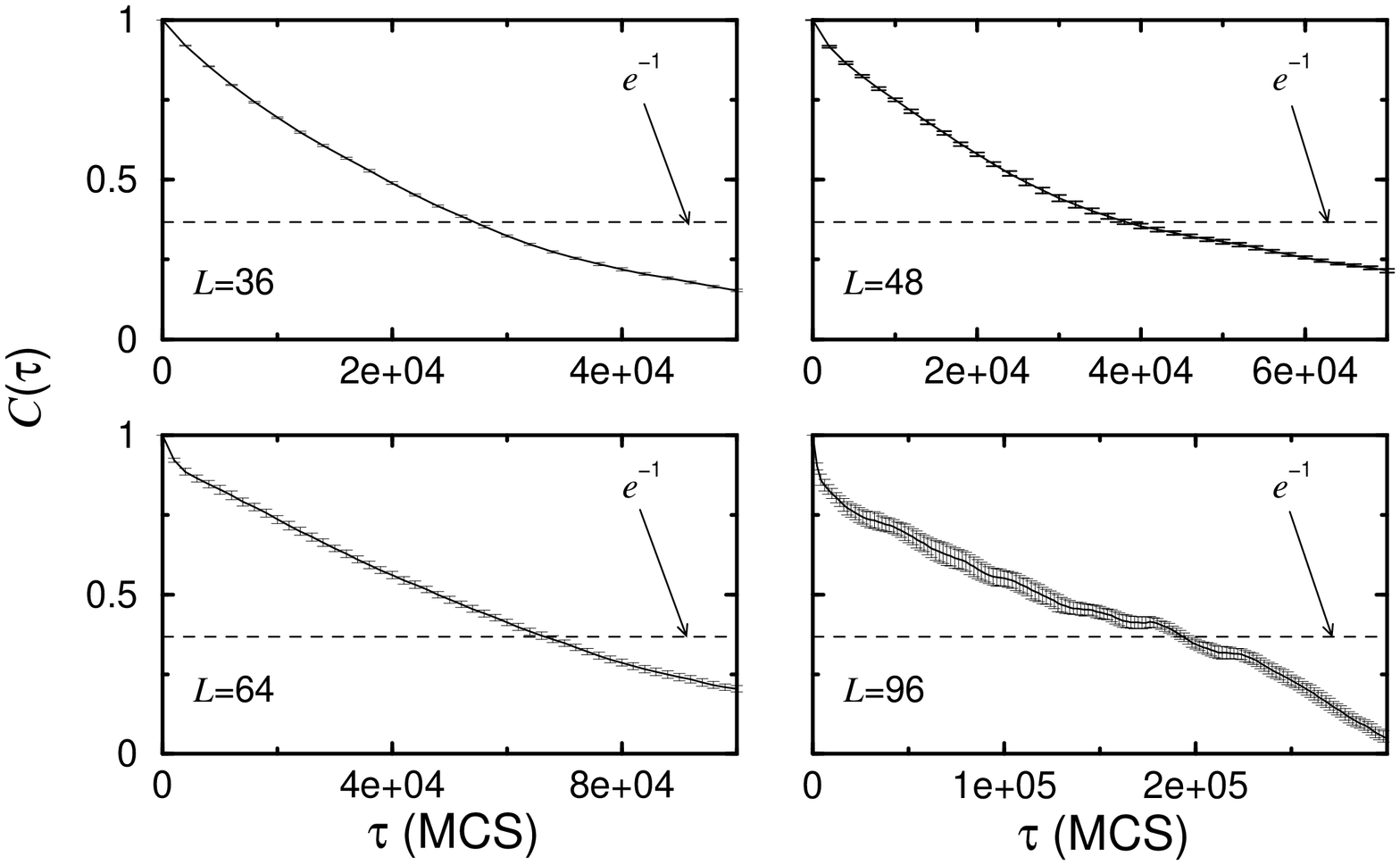}}}
\caption{ Average time correlation in the order parameter $m$ of the
  5-state Potts model. We fix the temperature $T=0.8515$ and vary the
  size of systems. The time delay $\tau_0$ corresponding to
  $C(\tau)=1/e$ is the characteristic time scale of
  interest.}
\label{appfig7}
\end{figure}

The order parameters which we obtain through Monte Carlo simulations
may be correlated in time. Such correlation may be stronger when the
system is in the vicinity of the critical point. As an example, here
we investigate the 5-state Potts model near the phase transition. We
fix the temperature $T=0.8515$ and vary the size of systems. To obtain
the time correlation function $C(\tau)$, for different sizes of
systems we fix the thermal averaging interval
$N_{\text{tw}}=2,000$. The order parameter $m_{i,\alpha}$ of the Potts
model depends on the spin value $\alpha$ and the spin index $i$. We
also fix $\alpha$ and for each original spin $i$ we calculate the time
correlation $C_i(\tau)$. We show the average $C(\tau)$ of all original
spins in Fig.~\ref{appfig7}. When the time decay $\tau$ is small
$C(\tau)$ decays exponentially, i.e., $C(\tau)\sim \exp[-\tau/\tau_0]$
where $\tau_0$ is the characteristic time scale. Comparing with the
results we show in Fig~2b, we find that the
distribution of the order parameter ratio $P(D)$ becomes symmetric
when the thermal averaging interval $N_{\text{tw}}$ is comparable or
larger than the characteristic time scale of the order parameter $m$.

\subsection{Multifractal detrended fluctuation analysis (MFDFA)}
\label{appendix7}
The multifractal detrended fluctuation analysis (MFDFA) is an
efficient and reliable method available to quantify the
multifractality in a non-stationary series.  For a series $\{x_i\}$
with the length $N$, its procedure is the following:

(1) Calculate the profile
$$Y(j)=\sum_{i=1}^j[x_i-\langle x \rangle], j=1,\ldots,N,$$ 
where $\langle \ldots \rangle$ is the mean of $\{x_i\}$.

(2) Divide $Y(j)$ into $N_n \equiv \text{int}(N/n)$ non-overlapping
parts of length $n$. The maximal value of $n$ should be smaller
than $N/4$ to avoid statistically unreliable results. We perform this
step from both the beginning and the end of the signal. Thus we
totally obtain $2N_n$ parts.

(3) For each part with index $\nu$ we calculate the local trend
$y_{\nu}^{\ell}(i)$ in this part by a least-square fit of the series
with a polynomial function, where $\ell$ is the order of the
polynomial function. We then subtract this local trend and determine
the variance:
$$F^2(\nu,n) \equiv \frac{1}{n}\sum_{i=1}^n \{ Y[(\nu-1)n+i]-y_{\nu}^{\ell}(i)\}^2,$$
for $\nu=1,\ldots,N_n$ and 
$$F^2(\nu,n) \equiv \frac{1}{n}\sum_{i=1}^n \{ Y[N-(\nu-N_n)n+i]-y_{\nu}^{\ell}(i)\}^2,$$
for $\nu=N_n+1,\ldots,2N_n.$

(4) Average over all parts we obtain the $r$-th order
fluctuation function:
$$ 
F_r(n) \equiv \left\{ \frac{1}{2N_n} \sum_{\nu=1}^{2N_n} [F^2(\nu,n)]^{r/2}
\right\}^{1/r}.
$$
For $r=0$ we choose 
$$
F_0(n) \equiv  \exp \left\{\frac{1}{4N_n}\sum_{\nu=1}^{2N_n} \ln [F^2(\nu,n)]
\right\}.
$$ 
(5) Determine the scaling of the fluctuation
function: $F_r(n) \sim n^{h(r)}$. If $h(r)$ is a constant, the signal
is monofractal; otherwise it is multifractal.

The singularity spectrum can be further calculated from: $f(\alpha)
=r[\alpha-h(r)]+1,$ where the singularity strength $\alpha=h(r) +
rh'(r)$. If the signal is monofractal, we find that
$\alpha=h(r)=\text{const}$ and $f(\alpha)=1$.

\subsection{Multifractal behaviour of the 5-state Potts model}
\label{appendix8}
\begin{figure}
\epsfysize=0.99\columnwidth{\rotatebox{-90}{\epsfbox{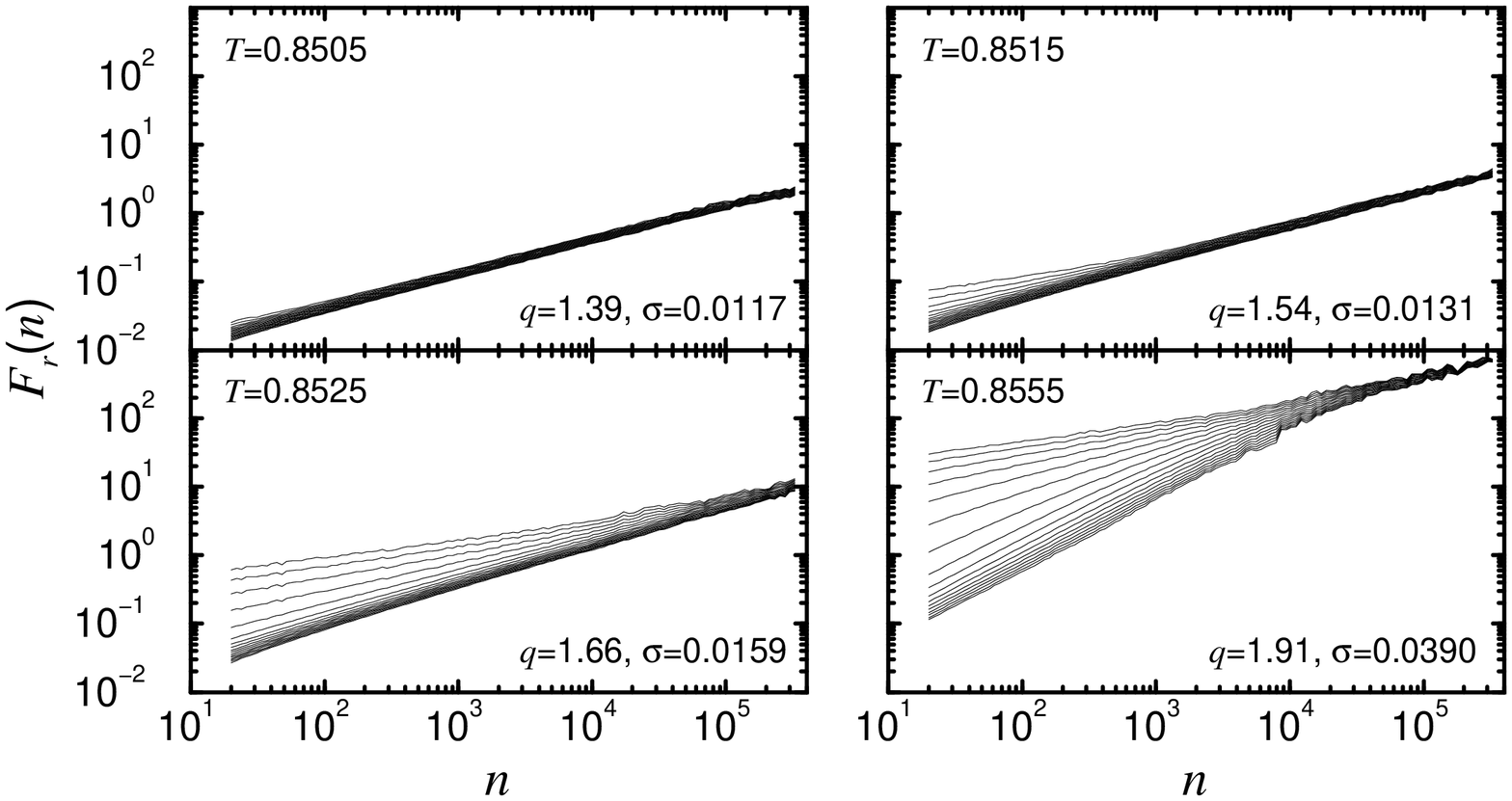}}}
\caption{Multifractal behaviour of the 5-state Potts model near the
  critical point. The system size is $L=48$ and we have applied the
  MFDFA. The data have been shuffled before the calculation and the
  length of the order parameter ratio is 2 million. We set the thermal
  averaging interval $N_{\text{tw}}=80,000$ at all temperatures. We
  show the fluctuation function $F_r(n)$ versus observing window $n$
  at different temperatures. Both the parameters $q$ and $\sigma$ are
  obtained from the $q$-Gaussian fit to the central part of $P(D)$.  }
\label{appfig8}
\end{figure}

As another example here we investigate the multifractal behaviour of
the 5-state Potts model in the vicinity of the critical point. We
set the thermal averaging interval $N_{\text{tw}}=80,000$ at all
temperatures. After obtaining the series of the order parameter ratio,
we first shuffle the data and then apply the MFDFA. The results of the
fluctuation function $F_r(n)$ versus the observing window $n$ are
shown in Fig.~\ref{appfig8}. We find that, below the critical point
(in the ordered phase) the system is monofractal since the slope of
$F_r(n)$ is identical for different values of $r$. When approaching
the critical point from the ordered phase, the slope of $F_r(n)$
starts to shift at small scales, indicating a multifractal behaviour
within these scales. Such region is broader when the system is closer
to the critical point. At the critical point we observe multifractal
behaviour at all scales we measure. When continuing increasing the
temperature, the system is in the disordered phase. The value of the
parameter $q$ from the $q$-Gaussian fit is increasing while the region
where the multifractal behaviour presents shrinks. Such behaviour is
similar to that of the Ising model. Nevertheless, for the Potts model
we find that the scale parameter $\sigma$ of the $q$-Gaussian fit is
increasing with increasing temperature and fixed $N_{\text{tw}}$,
while it is not sensitive to the temperature for the Ising model if
the measured $q$ is far below 2.


\subsection{Characteristics of the multifractal behaviour}
\label{appendix9}
\begin{figure}
\epsfysize=0.99\columnwidth{\rotatebox{-90}{\epsfbox{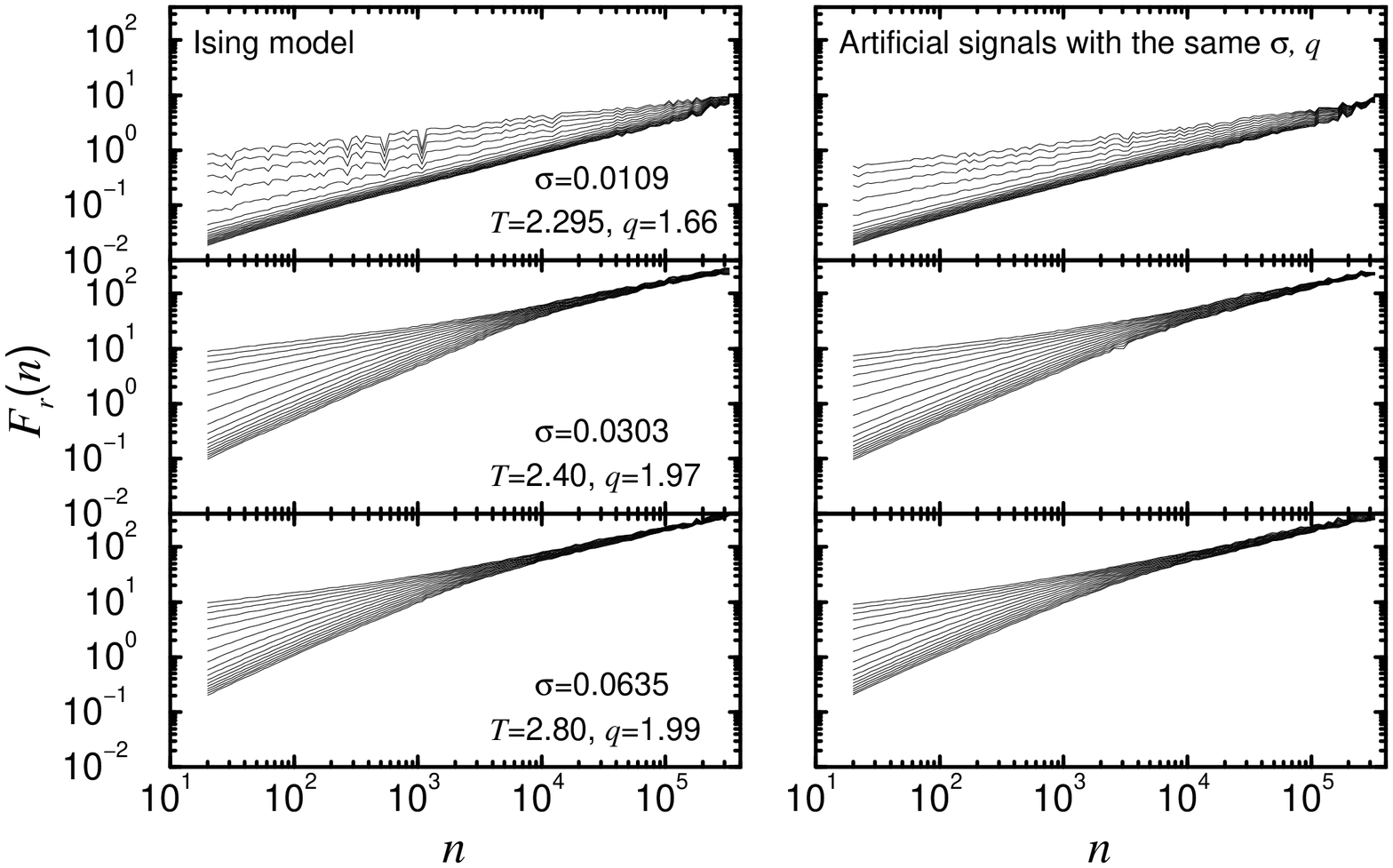}}}
\caption{Multifractal behaviour of the Ising model compared to that of
   $q$-Gaussian distributed artificial signals. The system size of
  the Ising model is $L=124$. We set the same parameters $\sigma$ and
  $q$ for the artificial signals and the model. The data length is 1.6
  million.}
\label{appfig9}
\end{figure}
Here we investigate the characteristics of the multifractal behaviour
we show in the paper. To do this, we generate some $q$-Gaussian
distributed artificial signals and we set the same parameters for the
artificial signals and for the testing model --- the Ising model. For the
model we have first shuffled the data thus eliminated the dynamics in
signals. As shown in Fig.~\ref{appfig9} we find that the results
are almost identical for the artificial signals and the model when the
parameters $q$ and $\sigma$ are the same. This verifies that the
multifractal behaviour we present in the paper is not a system
specific behaviour but should be applicable to other physical systems.

\begin{figure}[h!]
\epsfysize=0.99\columnwidth{\rotatebox{-90}{\epsfbox{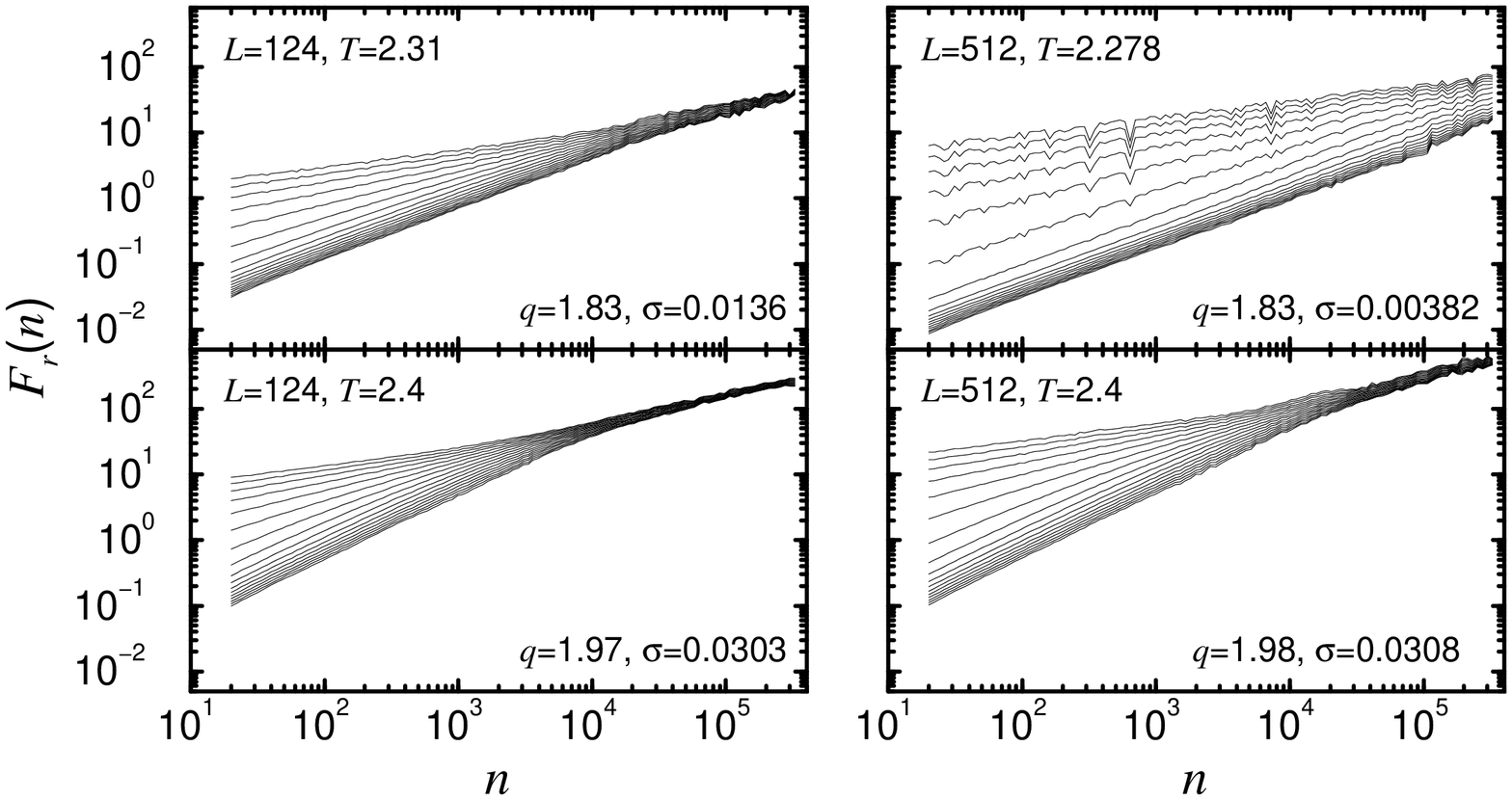}}}
\caption{System size dependence of the multifractal behaviour of the
  Ising model. The data length is 1.6 million.}
\label{appfig10}
\end{figure}

We then investigate how the multifractal behaviour depends on the size
of a specific system. Such dependence should be manifested from the
values of two parameters $q$ and $\sigma$. To see this, we investigate
the Ising model with different sizes $L=124$ and $L=512$. We first
shuffle the data and eliminate the dynamics in signals. We then apply
the MFDFA and find that with the same parameter $q$, the difference in
the observed multifractal behaviour is owing to different values of
the scale parameter $\sigma$ (see Fig.~\ref{appfig10}). When $q$ is
fixed, the larger system with smaller value of $\sigma$ shows
multifractal behaviour in broader region than the smaller system
shows.

\end{document}